\documentclass[aps,superscriptaddress,nofootinbib,12pt,tightenlines]{revtex4}



\usepackage{natbib}
\usepackage{graphicx}
\usepackage{dcolumn}
\begin{document}
% You should use BibTeX and apsrev.bst for references
%\bibliographystyle{apsrev}

% Use the \preprint command to place your local institutional report
% number on the title page in preprint mode.
% Multiple \preprint commands are allowed.
%\preprint{}

%Title of paper
\title{There's more to volatility than volume}
% Optional argument for running titles on pages
%\title[]{}

% repeat the \author .. \affiliation  etc. as needed
% \email, \thanks, \homepage, \altaffiliation all apply to the current
% author. Explanatory text should go in the []'s, actual e-mail
% address or url should go in the {}'s for \email and \homepage.
% Please use the appropriate macro for the type of information

% \affiliation command applies to all authors since the last
% \affiliation command. The \affiliation command should follow the

\author{L\'aszl\'o Gillemot}
%\email[]{jdf@santafe.edu}
\affiliation{Santa Fe Institute, 1399 Hyde Park Road, Santa Fe, NM 87501}
\affiliation{Budapest University of Technology and Economics,
H-1111 Budapest, Budafoki \'ut 8, Hungary}

\author{J. Doyne Farmer}
%\email[]{jdf@santafe.edu}
\affiliation{Santa Fe Institute, 1399 Hyde Park Road, Santa Fe, NM 87501}

\author{Fabrizio Lillo}
%\email[]{lillo@santafe.edu}
\affiliation{Santa Fe Institute, 1399 Hyde Park Road, Santa Fe, NM 87501}
\affiliation{INFM-CNR Unit\`a di Palermo and Dipartimento di Fisica e Tecnologie Relative, 
Universit\`a di Palermo, Viale delle Scienze, I-90128 Palermo, Italy.}

\begin{abstract}
It is widely believed that fluctuations in transaction volume, as reflected in the number of transactions and to a lesser extent their size, are the main cause of clustered volatility.  Under this view bursts of rapid or slow price diffusion reflect bursts of frequent or less frequent trading, which cause both clustered volatility and heavy tails in price returns.  We investigate this hypothesis using tick by tick data from the New York and London Stock Exchanges and show that only a small fraction of volatility fluctuations are explained in this manner. Clustered volatility is still very strong even if price changes are recorded on intervals in which the total transaction volume or number of transactions is held constant. In addition the distribution of price returns conditioned on volume or transaction frequency being held constant is similar to that in real time, making it clear that neither of these are the principal cause of heavy tails in price returns.  We analyze recent results of Ane and Geman (2000) and Gabaix et al. (2003), and discuss the reasons why their conclusions differ from ours.  Based on a cross-sectional analysis we show that the long-memory of volatility is dominated by factors other than transaction frequency or total trading volume.
\end{abstract}

% insert suggested PACS numbers in braces on next line
%\pacs{}
% insert suggested keywords - APS authors don't need to do this
%\keywords{}

%\maketitle must follow title, authors, abstract, \pacs, and \keywords
\maketitle
\tableofcontents

\section{Introduction}
The origin of heavy tails and clustered volatility in price fluctuations (Mandelbrot 1963) is an important problem in financial economics\footnote{The observation that price fluctuations display heavy tails and clustered volatility is quite old.  In his 1963 paper Mandelbrot notes that prior references to observations of heavy tails in prices date back to 1915.  He also briefly discusses clustered volatility, which he says was originally pointed out to him by Hendrik Houthakker.  The modern upsurge of interest in clustered volatility stems from the work of Engle (1982)}.  Heavy tails means that large price fluctuations are much more common than they would be for a normal distribution, and clustered volatility means that the size of price fluctuations has strong positive autocorrelations, so that there are periods of large or small price change. Understanding these phenomena has considerable practical importance for risk control and option pricing.  Although the cause is still debated, the view has become increasingly widespread that in an immediate sense both of these features of prices can be explained by fluctuations in volume, particularly as reflected by the number of transactions. In this paper we show that while fluctuations in volume or number of transactions do indeed affect prices, they do not play the dominant role in determining either clustered volatility or heavy tails.

The model that is now widely believed to explain the statistical properties of prices has its origins in a proposal of Mandelbrot and Taylor (1967) that was developed by Clark (1973). Mandelbrot and Taylor proposed that prices could be modeled as a subordinated random process $Y(t) = X(\tau(t))$, where $Y$ is the random process generating returns, $X$ is Brownian motion and $\tau(t)$ is a stochastic time clock whose increments are IID and uncorrelated with $X$. Clark hypothesized that the time clock $\tau(t)$ is the cumulative trading volume in time $t$. Since then several empirical studies have demonstrated a strong correlation between volatility Ð- measured as absolute or squared price changes Ð- and volume (Tauchen and Pitts 1983, Karpoff 1987, Gerity and Mulherin 1989, Stephan and Whaley 1990, Gallant, Rossi and Tauchen 1992). More recently evidence has accumulated that the number of transactions is more important than their size (Easley and OÕHara 1992, Jones, Kaul and Lipsom 1994, Ane and Geman 2000).  We show that volatility is still very strong even if price movements are recorded at intervals containing an equal number of transactions, and that the volatility observed in this way is highly correlated with volatility measured in real time.  In contrast, if we shuffle the order of events, but control for the number of transactions so that it matches the number of transactions in real time, we observe a much smaller correlation to real time volatility.  We interpret this to mean that the number of transactions is less important than other factors.

Several studies have shown that the distribution of price fluctuations can be transformed to a normal distribution under a suitable rescaling transformation (Ane and Geman 2000, Plerou et al. 2000, Andersen et al. 2003).  There are important differences between the way this is done in these studies.  Ane and Geman claim that the price distribution can be made normal based on a transformation that depends only on the transaction frequency.  The other studies, in contrast, use normalizations that also depend on the price movements themselves.  Plerou et al. divide the price movement by the product of the square root of the transaction frequency and a measure of individual price movements, and Andersen et al. divide by the standard deviation of price movements.  We find that, contrary to Ane and Geman, it is not sufficient to normalize by the transaction frequency, and that in fact in many cases this hardly alters the distribution of returns at all.  A related theory is due to Gabaix et al. (2003), who have proposed that the distribution of size of large price changes can be explained based on the distribution of volume and the volume dependence of the price impact of trades.  We examine this hypothesis and show that when prices are sampled to hold the volume constant the distribution of price changes, and in particular its tail behavior, is similar to that in real time.  We present some new tests of their theory that give insight into why it fails.  Finally we study the long-memory properties of volatility and show that neither volume nor number of transactions can be the principal causes of long-memory behavior.

Note that when we say a factor ``causes" volatility, we are talking about proximate cause as opposed to ultimate cause.  For example, changes in transaction frequency may cause changes in volatility, but for the purposes of this paper we do not worry about what causes changes in transaction frequency.  It may be the case that causality also flows in the other direction, e.g. that volatility also causes transaction frequency.  All we test here are correlations.  In saying that transaction frequency or trading volume cannot be a dominant cause of volatility, we are making the assumption that if the two are not strongly correlated there cannot be a strong causal connection in either direction\footnote{We recognize that if the relationship between two variables is highly nonlinear it is possible to have a strong causal connection even if the variables are uncorrelated.  Thus, in claiming a lack of causality, we are assuming that nonlinearities in the relationship between the variables are small enough that correlation is sufficient.}.

The paper is organized as follows: In Section II we describe the data sets and the market structure of the LSE and NYSE.  In Section III we define volume and transaction time and discuss how volatility can be measured in each of them, and discuss the construction of surrogate data series in which the data are scrambled but either volume or number of transactions is held constant.  We then use both of these tools to study the relationship between volume and number of transactions and volatility.  In Section IV we discuss the distributional properties of returns and explain why our conclusions differ from those of Ane and Geman and Gabaix et al.  In Section V we study the long-memory properties of volatility, and show that while fluctuations in volume or number of transactions are capable of producing long-memory, other factors are more important.  Finally in the conclusions we summarize and discuss what we believe are the most important proximate causes of volatility.

\section{Data and market structure}

This study is based on data from two markets, the New York Stock Exchange (NYSE) and the London Stock Exchange (LSE).   For the NYSE we study two different subperiods, the 626 trading day period from Jan 1, 1995 to Jun 23, 1997, labeled NYSE1, and the 734 trading day period from Jan 29, 2001 to December 31, 2003, labeled NYSE2. These two periods were chosen to highlight possible effects of tick size. During the first period the average price of the stocks in our sample was $73.8\$$ and the tick size was an eighth of a dollar, whereas during the second period the average price was $48.2\$$ and the tick size was a penny. The tick size in the second period is thus substantially smaller than that in the first, both in absolute and relative terms.  For each data set we have chosen 20 stocks with high capitalizations.  The tickers of the 20 stocks in the NYSE1 set are AHP, AIG, BMY, CHV, DD, GE, GTE, HWP, IBM, JNJ, KO, MO, MOB, MRK, PEP, PFE, PG, T, WMT, and XON, and the tickers of the stocks in the NYSE2 set are AIG, BA, BMY, DD, DIS, G, GE, IBM, JNJ, KO, LLY, MO, MRK, MWD, PEP, PFE, PG, T, WMT, and XOM. There are a total of about 7 million transactions in the NYSE1 data set and 36 million in the NYSE2 data set. 

For the LSE we study the period from May 2000 to December 2002, which contains 675 trading days.
For LSE stocks there has been no overall change in tick size during this period. The average price of a stock in the sample is about 500 pence, though the price varies considerably - it is as low as 50 pence and as high as 2500 pence. The tick size depends on the stock and the time period, ranging from a fourth of a pence to a pence.  As for the NYSE we chose stocks that are heavily traded.  The tickers of the 20 stocks in the LSE sample are AZN, REED, HSBA, LLOY, ULVR, RTR, PRU, BSY, RIO, ANL, PSON, TSCO, AVZ, BLT, SBRY, CNA, RB., BASS, LGEN, and ISYS.  In aggregate there are a total of 5.7 million transactions during this period, ranging from 497 thousand for HSBA to 181 thousand for RB.

The NYSE and the LSE have dual market structures consisting of a centralized limit order book market and a decentralized bilateral exchange.  The centralized limit order book market is called the downstairs market in New York and the on-book market in London, and the decentralized bilateral exchange is called the upstairs market in New York and the off-book market in London. While the corresponding components of each market are generally similar between London and New York, there are several important differences. 

The downstairs market of the NYSE operates through a specialist system. The specialist is given monopoly privileges for a given stock in return for committing to regulatory responsibilities. The specialist keeps the limit order book, which contains limit orders with quotes to buy or sell at specific prices. As orders arrive they are aggregated, and every couple of minutes orders are matched and the market is cleared. Trading orders are matched based on order of arrival and price priority. During the time of our study market participants were allowed to see the limit order book only if they were physically present in the market and only with the permission of the specialist.  The specialist is allowed to trade for his own account, but also has regulatory duties to maintain an orderly market by making quotes to ensure that prices do not make large jumps.  Although a given specialist may handle more than one stock, there are many specialists so that two stocks chosen at random are likely to have different specialists. 

The upstairs market of the NYSE, in contrast, is a bilateral exchange. Participants gather informally or interact via the telephone. Trades are arranged privately and are made public only after they have already taken place (Hasbrouck, Sofianos and Sosebee, 1993). 

The London Stock Exchange also consists of two markets (London Stock Exchange, \citeyear{LSE04}). The on-book market (SETS) is similar to the downstairs market and the off-book market (SEAQ) is similar to the upstairs market.  In 1999 $57\%$ of transactions of LSE stocks occurred in the on-book exchange and in 2002 this number rose to $62\%$.  One important difference between the two markets is for the on-book exchange quotations are public and are published without any delay.  In contrast, since transactions in the off-book market are arranged privately, there are no published quotes.  Transaction prices in the off-book market are published only after they are completed.  Most (but not all) of our analysis for the LSE is based on the on-book market. 

The on-book market is a fully automated electronic exchange. Market participants have the ability to view the entire limit order book at any instant, and to place trading orders and have them entered into the book, executed, or canceled almost instantaneously. Though prices and quotes are completely transparent, the identities of parties placing orders are kept confidential, even to the two counter parties in a transaction. The trading day begins with an opening auction. There is a period of 10 minutes leading up to the opening auction when orders may be placed or cancelled without transactions taking place, and without any information about volumes or prices. The market is then cleared and for the remainder of the day (except for rare exceptions) there is a continuous auction, in which orders and cancellations are entered asynchronously and result in immediate action. For this study we ignore the opening and closing auctions and analyze only orders placed during the continuous auction. 

The off-book market (SEAQ) is very similar to the upstairs market in the NYSE. The main difference is that there is no physical gathering place, so transactions are arranged entirely via telephone. There are also several other minor differences, e.g. regarding the types of allowed transactions and the times when trades need to be reported. For example, a trade in the off-book market can be an ordinary trade, with a fixed price that is agreed on when the trade is initiated, or it can be a VWAP trade, where the price depends on the volume weighted average price, which is not known at the time the trade is arranged. In the latter case the trade does not need to be reported until the trade is completed. In such cases delays of a day or more between initiation and reporting are not uncommon. 

For the NYSE we use the Trades and Quotes (TAQ) data set. This contains the prices, times, and volume of all transactions, as well as the best quoted prices to buy or sell and the number of shares available at the best quotes at the times when these quotes change. Transaction prices from the upstairs and downstairs markets are mixed together and it is not possible to separate them. Our analysis in the NYSE is based on transaction prices, though we have also experimented with using the average price of the best quotes.  For the 15 minute time scale and the heavily traded stocks that we study in this paper this does not seem to make very much difference. 

The data set for the LSE contains every order and cancellation for the on-book exchange, and a record of all transactions in both the on-book and off-book exchanges. We are able to separate the on-book and off-book transactions. The on-book transactions have the advantage that the timing of each trade is known to the nearest second, whereas as already mentioned, the off-book transactions may have reporting delays. Also the on-book transactions are recorded by a computer, whereas the off-book transactions depend on human records and are more error prone. Because of these problems, unless otherwise noted, we use only the downstairs prices. Our analysis for the LSE is based entirely on mid-quote prices in the on-book market, defined as the average of the best quote to buy and the best quote to sell.

To simplify our analysis we paste the time series for each day together, omitting any price changes that occur outside of the period of our analysis, i.e. when the market is closed.  For the NYSE we omit the first few and last few trades, but otherwise analyze the entire trading day, whereas for the LSE we omit the first and last half hour of each trading day.  We don't find that this makes very much difference -- except as noted in Section~\ref{GabaixSection}, our results are essentially unchanged if we include the entire trading day.

\section{Volatility under alternative time clocks}

\subsection{Alternative time clocks}

A stochastic time clock $\tau(t)$ can speed up or slow down a random process and thereby alter the distribution or autocorrelations of its discrete increments.  When the increments of a stochastic time clock are IID it is called a subordinator, and the process $Y(t) = X(\tau(t))$ is called a subordinated random process.  The use of stochastic time clocks in economics originated in the business literature (Burns and Mitchell 1946), and was suggested in finance by Mandelbrot and Taylor (1967) and developed by Clark (1973).  {\it Transaction time} $\tau_\theta$ is defined as  
\begin{equation}
\tau_\theta(t_i) = \tau_\theta(t_{i-1}) + 1\nonumber,
\end{equation}
where $t_i$ is the time when transaction $i$ occurs.  Another alternative is the {\it volume time} $\tau_v$, defined as 
\begin{equation}
\tau_v(t_i)=\tau_v(t_{i-1})+V_i, \nonumber
\end{equation}
where $V_i$ is the volume of transaction $i$.

The use of transaction time as opposed to volume time is motivated by the observation that the average price change caused by a transaction, which is also called the average market impact, increases slowly with transaction size.  This suggests that the number of transactions is more important than their size (Hasbrouck 1991, Easley and OÕHara 1992, Jones, Kaul and Lipson 1994, Lillo, Farmer and Mantegna 2003). 

\begin{figure}[ptb]
   \begin{center}
     \includegraphics[scale=0.6]{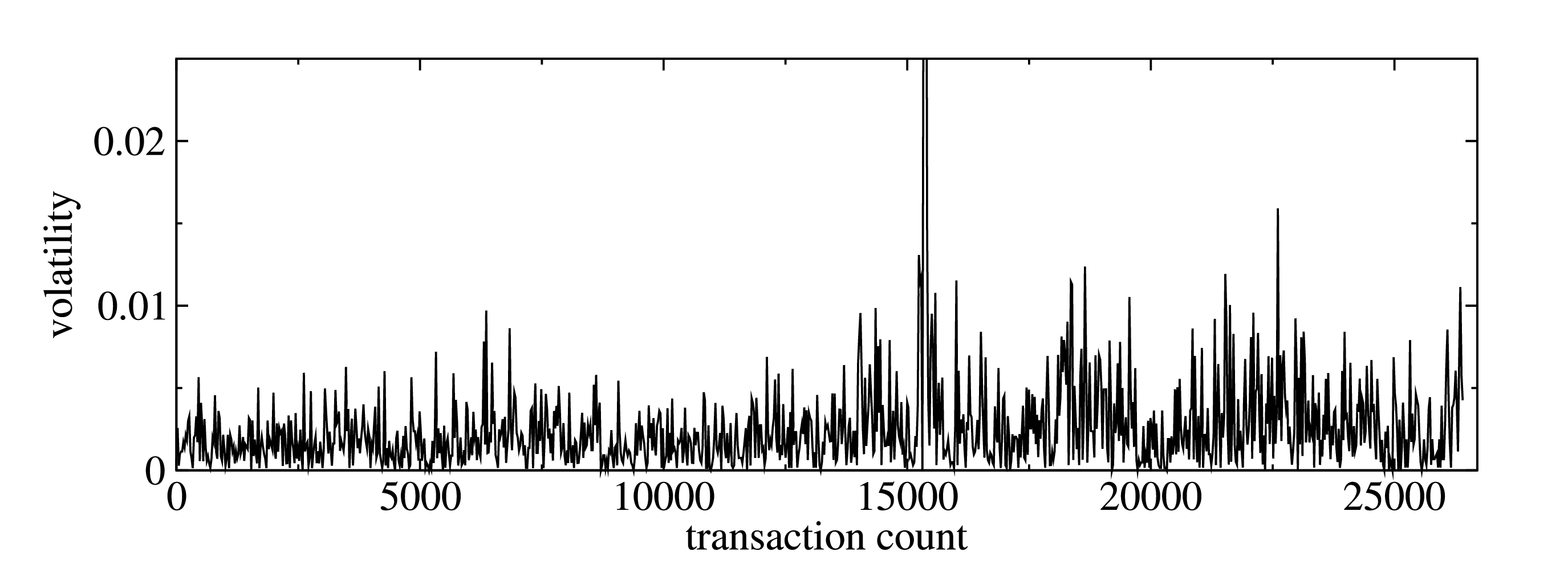}
     \caption{ Transaction time volatility for the stock AZN, which is defined as the absolute value of the price change over an interval of $24$ transactions. The resulting intervals are on average about fifteen minutes long. The series runs from August 1, 2001 to Sept. 27, 2001 and includes 26,400 transactions. }
    \label{volatilityEventVsReal}
  \end{center}
\end{figure}

\subsection{Volatility in transaction time}

The main results of this paper were motivated by the observation that clustered volatility remains strong even when the price is observed in transaction time. Letting $t_i$ denote the time of the $i^{th}$ transaction, we define the {\it transaction time volatility}\footnote{We use the absolute value of end-to-end price changes rather than the standard deviation to avoid sampling problems when comparing different time clocks. We use the midquote price for the LSE to avoid problems with mean reversion of prices due to bid-ask bounce. We find that on the time scales we study here this makes very little difference. For convenience in the NYSE data series we use the transaction price.} for interval $(t_{i-K}, t_i)$ as $\nu_\theta(t_i, K) = |p(t_i)-p(t_{i-K})|$, where $p(t_i)$ is the logarithm of the midquote price at time $t_i$. Here and in all the other definitions of volatility we use always non-overlapping intervals, $p$ will refer to the logarithm of the price, and when we say ``return" we always mean log returns, i.e. $r(t) = p(t) - p(t-\Delta t)$.  A transaction time volatility series for the LSE stock Astrazeneca with $K = 24$ transactions is shown in Figure 1.  For $K = 24$ the average duration is 15 minutes, though there are intervals as short as 14 seconds and as long as 159 minutes\footnote{Remember that we are omitting all price changes outside of trading hours.}. In Figure 1 there are obvious periods of high and low volatility, making it clear that volatility is strongly clustered, even through we are holding the number of transactions in each interval constant. 

For comparison in Figure 2(a) we plot the real time volatility $\nu(t, \Delta t) = |p(t) - p(t - \Delta t)|$ based on $\Delta t =$ fifteen minutes, over roughly the same period shown in Figure 1.  In Figure 2(b) we synchronize the transaction time volatility with real time by plotting it against real time rather than transaction time. It is clear from comparing panel (a) and panel (b) that there is a strong contemporary correlation between real time and transaction time volatility. 

As a further point of comparison in panel (c) we randomly shuffle transactions.  We do this so that we match the number of transactions in each real time interval, while preserving the unconditional distribution of returns but destroying any temporal correlations. Let the (log) return corresponding to transaction $i$ be defined as $r(t_i) = p(t_i)-p(t_{i-1})$. The shuffled transaction price series $\tilde p(t_i)$ is created by setting $\tilde r(t_i) = r(t_j)$, where $r(t_j)$ is drawn randomly from the entire time series without replacement.  We then aggregate the individual returns to create a surrogate price series $\tilde p(t_i) =\sum_{k=1}^i \tilde r(t_k) + p(t_0)$. We define the {\it shuffled transaction real time volatility} as $\tilde\nu_\theta(t, \Delta t) = | \tilde p(t)- \tilde p(t - \Delta t)|$. The name emphasizes that even though transactions are shuffled to create a surrogate price series, the samples are taken at uniform intervals of real time and the number of transactions matches that in real time. The shuffled transaction real time volatility series shown in Figure 2(c) is sampled at fifteen minute intervals, just as for real time. The resulting series still shows some clustered volatility, but the variations are smaller and the periods of large and small volatility do not match the real time volatility as well. This indicates that the number of transactions is less important in determining volatility than other factors, which persist even when the number of transactions is held constant. 

\begin{figure}[ptb]
   \begin{center}
     \includegraphics[scale=0.6]{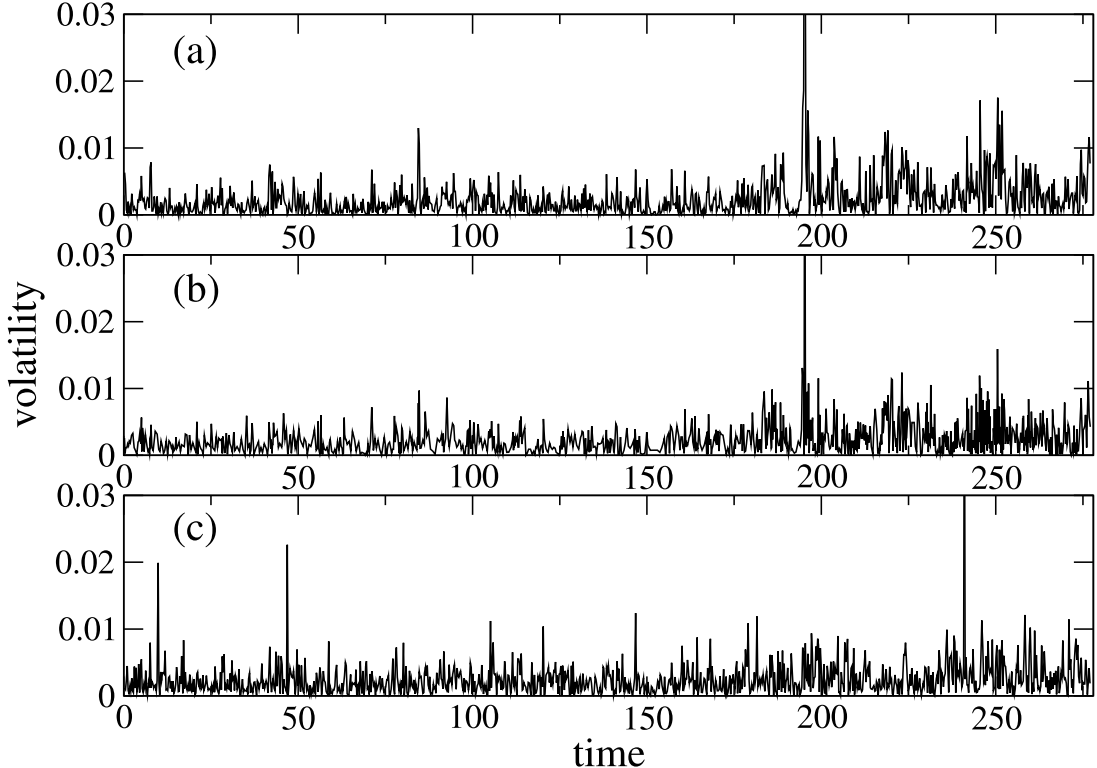}
     \caption{ A comparison of three different volatility series for the LSE stock Astrazeneca. In panel (a) the volatility is measured over real time intervals of fifteen minutes. In panel (b) the transaction time volatility is measured at intervals of 24 transactions, which roughly correspond to fifteen minutes, and then plotted as a function of real time. Panel (c) shows the volatility of a price series constructed by preserving the number of transactions in each $15$ minute interval, but drawing single transaction returns randomly from the whole time series and using them to construct a surrogate price series. The fact that the volatilities in (a) and (b) agree more closely than those in (a) and (c) suggests that fluctuations in transaction frequency plays a smaller role in determining volatility than other factors. Time is measured in hours and the the period is the same as in Fig.~1}
    \label{volatilityEventVsReal}
  \end{center}
\end{figure}

\begin{figure}[ptb]
   \begin{center}
     \includegraphics[scale=0.5]{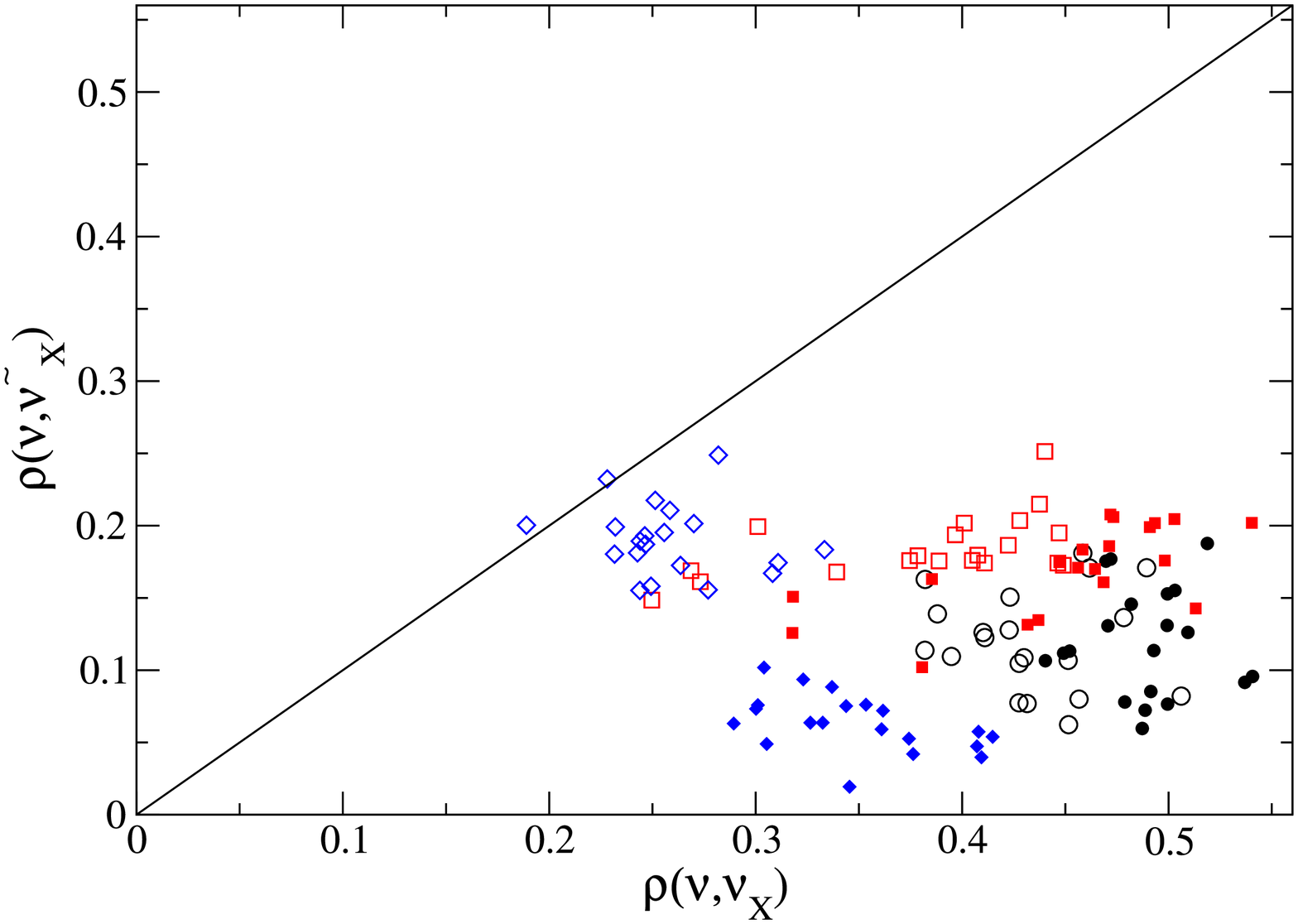}
     \caption{An illustration of the relative influence of number of transactions and volume on volatility. The correlation between real time volatilty $\nu$ and various alternative volatilities $\nu_x$ is plotted against the correlation with the corresponding volatility $\tilde{\nu}_x$ when the order of events is shuffled.  Each mark corresponds to a particular stock, data set, and time clock.  The solid marks are for the transaction related quantities, i.e. $\nu_x = \nu_\theta$ and $\tilde{\nu}_x = \tilde{\nu}_\theta$, and the open marks are for volume related quantities, i.e. $\nu_x = \nu_v$ and $\tilde{\nu}_x = \tilde{\nu}_v$.   Thus the horizontal axis represents the correlation with volume or transaction time, and the vertical axis represents the correlation with shuffled volume or shuffled transaction real time.   Stocks in the LSE are labeled with black circles, in the NYSE1 (1/8 tick size) with red squares, and the NYSE2 (penny tick size) with blue diamonds. Almost all the data lie well below the identity line, demonstrating that volume or the number of transactions has a relatively small effect on volatility -- most of the information is present when these are held constant.}
    \label{volatilityEventVsReal}
  \end{center}
\end{figure}

\subsection{Correlations between transaction time and real time}

To measure this the influence of number of transactions on prices more quantitatively we compare the correlations between the volatilities of these series. To deal with the problem that the sampling intervals in real time and transaction time are different we first convert both the real time and transaction time volatility series, $\nu(t_i, K)$ and $\nu_{\theta}(t_i, K)$  to continuous time using linear interpolation within each interval $(t_{i-K}, t_i)$. This allows us to compute $\rho(\nu, \nu_\theta)$, the correlation between real time and transaction time volatilities as they were continuous functions. 
For comparison we also compute $\rho(\nu, \tilde\nu_\theta)$, the correlation between the real time and the shuffled transaction real time volatility series. This is done for each stock in all three data sets. The results are plotted in Figure 3 and summarized in Table I. In every case $\rho(\nu, \tilde{\nu}_\theta)$ is significantly smaller than $\rho(\nu, \nu_\theta)$. The correlations between real time volatility and transaction time volatility are between $35\%$ and $49\%$, whereas the correlation with shuffled transaction real time volatility is between $6\%$ and $17\%$. The ratio is typically about a factor of four. This demonstrates that fluctuations in transaction rates account for only a small portion of the variation in volatility. This quantitatively confirms what the eye sees in Figure 2 -- the principal influences on volatility are still present even if the number of transactions are held constant.  The way in which prices respond to transactions is much more important than the number of transactions. 

\begin{table}
\caption{Average correlations and ratios of average correlations between real time and alternate volatilities. Correlations are given as percentages, and error bars are standard deviations for the data shown in Figure 3. All the correlations are against the real time volatility $\nu$. $E[x]$ denotes the sample mean of $x$ averaging across stocks. The first column is the correlation with transaction time volatility, the second column with shuffled transaction real time volatility, the third with volume time volatility, and the fourth with shuffled volume real time volatility. The fifth column is the ratio of the first and second columns and the sixth column is the ratio of the third and fourth columns. The fact that the values in the last column are all greater than one shows that fluctuations in transaction number and volume are minority effects on real time volatility. }
\begin{tabular}{lllllll}
\tableline
data set & $E[\rho(\nu, \nu_\theta)]$ & $E[\rho(\nu, \tilde\nu_\theta)]$ &$ E[\rho(\nu, \nu_v)]$ &  $E[\rho(\nu, \tilde\nu_v)]$ & $\frac{E[\rho(\nu,\nu_\theta)]}{E[\rho(\nu,\tilde\nu_\theta)]}$ & $\frac{E[\rho(\nu,\nu_v)]}{E[\rho(\nu,\tilde\nu_v)]}$ \\
\tableline
LSE & $49 \pm 3$ & $12 \pm 4$ & $43 \pm 3$ & $12 \pm 3$ & $4.1$ & $3.6$\\
NYSE1 &$45 \pm 6$ & $17 \pm 3$ & $38 \pm 6$ & $18 \pm 2$ & $2.6$ & $2.1$\\ 
NYSE2& $35 \pm 4$ & $6 \pm 2$& $26 \pm 3$ & $19 \pm 2$ & $5.8$ & $1.4$\\ 
\tableline
\end{tabular}
\label{correlationTable}
\end{table}

\subsection{Correlations between volume time and real time\label{volumeTime}}

 We can test the importance of volume in relation to volatility in a similar manner. To do this we create a volatility series $\nu_v$ sampled in volume time. Creating such a series requires sampling the price at intervals of constant volume.  This is complicated by the fact that transactions occur at discrete time intervals and have highly variable volume, which causes volume time $\tau_v(t)$ to increase in discrete and irregular steps. This makes it impossible to sample volume at perfectly regular intervals. We have explored several methods for sampling at approximately regular intervals, and have found that all the sampling procedures we tested yielded essentially the same results. For the results we present here we choose the simple method of adjusting the sampling interval upward to the next transaction. 
 
To make it completely clear how our results are obtained, we now describe our algorithm for approximate construction of equal volume intervals in more detail.  To ensure that the number of intervals in the volume time series is roughly equal to the number in the real time series, we choose a volume time sampling interval $\Delta V$ so that $\Delta V \approx \sum_i V_i/n$, where $n$ is the number of real time intervals and $\sum_i V_i$ is the total transaction volume for the entire time series.  We construct sampling intervals by beginning at the left boundary of each interval.  Assume the first transaction in the interval is transaction $i$.  We examine transactions $k = i-K, . . . , i$,  aggregating the volume $V_k$ of each transaction and increasing $K$ incrementally until $\sum_{k=i-K}^{i}  V_k > \Delta V$ . If the resulting time interval $(t_{i-K}, t_i)$ crosses a daily boundary we discard it. The {\it volume time volatility} is then defined to be $\nu_v(t_i, \Delta V ) = |p(t_i)-p(t_{i-K})|$, i.e. the absolute price change across the smallest interval whose volume is greater than $\Delta V$. We have also tried other more exact procedures, such as defining the price at the right boundary of the interval by interpolating between $p(t_{i-K})$ and $p(t_{i-K+1})$, but they give essentially the same results. 
 
In a manner analogous to the definition of shuffled transaction real time volatiliy, we can also define the {\it shuffled volume real time volatility} $\tilde\nu_v$.   This is constructed as a point of comparison to match the volume in each interval of the real time series but otherwise destroy any time correlations.  Our algorithm for computing the shuffled real time volatility is as follows:   We begin at a time $t$ corresponding to the beginning of a real time interval $(t - \Delta t, t)$ that we want to match, which contains transaction volume $V_{t-\Delta t, t}$.  We then randomly sample transactions $i$ from the entire time series and aggregate their volumes $V_i$, increasing $K$ until $\sum_{i=1}^K V_i > V_{t-\Delta t, t}$. We define the aggregate return for this interval as $\tilde{R}_{t-\Delta t, t} = \sum_{i=1}^K \tilde{r}(t_i)$, where $\tilde{r}(t_i)$ is the (log) return associated with each transaction. The shuffled volume real time volatility is $\tilde\nu_v(t,\Delta t) = |\tilde{R}_{t-\Delta t, t}|$. By comparing the volatility defined in this way to the real time volatility we can see how important volume is in comparison to other factors.

As with transaction time we can compute the correlations between volume time and real time volatility, $\rho(\nu, \nu_v)$, and the correlation between shuffled volume real time volatility and real time volatility, $\rho(\nu, \tilde{\nu}_v)$.  The results are summarized by the open marks in Figure 3 and columns three, four and six of Table I.  In Figure 3 we see that $\rho(\nu, \nu_v)$ is larger than $\rho(\nu, \tilde{\nu}_v)$ in almost every case.  The ratio $E[\rho(\nu, \nu_v)/\rho(\nu, \tilde{\nu}_v)]$ measures the average relative importance of factors that are independent of volume to factors that depend on volume for each of our three data sets.  In column 6 of Table I  these average ratios range from $1.4 - 3.6$.   As before, other factors have a bigger correlation to volatility than fluctuations in volume.  Thus neither volume fluctuations nor transaction rate fluctuations can account for the majority of variation in volatility. 

\section{Distributional properties of returns}

\subsection{Holding volume or transactions constant}

In this section we examine whether fluctuations in either transaction frequency or total trading volume play a major role in determining the distribution of returns. In Figure 4 we compare the cumulative distribution of absolute returns in real time to the distribution of absolute returns in transaction time and volume time for a couple of representative stocks\footnote{We study absolute returns throughout because the positive and negative tails are essentially the same, and because our primary interest is in volatility.}.  For London, where we can separate volume in the on-book and off-book markets, we compare both to total volume time (on book and off-book) and to volume time based on the on-book volume only. For the NYSE data we cannot separate these, and volume time is by definition total volume time.  We see that the differences between real time and either volume or transaction time are not large, indicating that transaction frequency and volume have a relatively small effect on the distribution of prices.

\begin{figure}[ptb]
   \begin{center}

     %\vspace{+0.05in}
     \includegraphics[height=1.9in, width=3.2in]{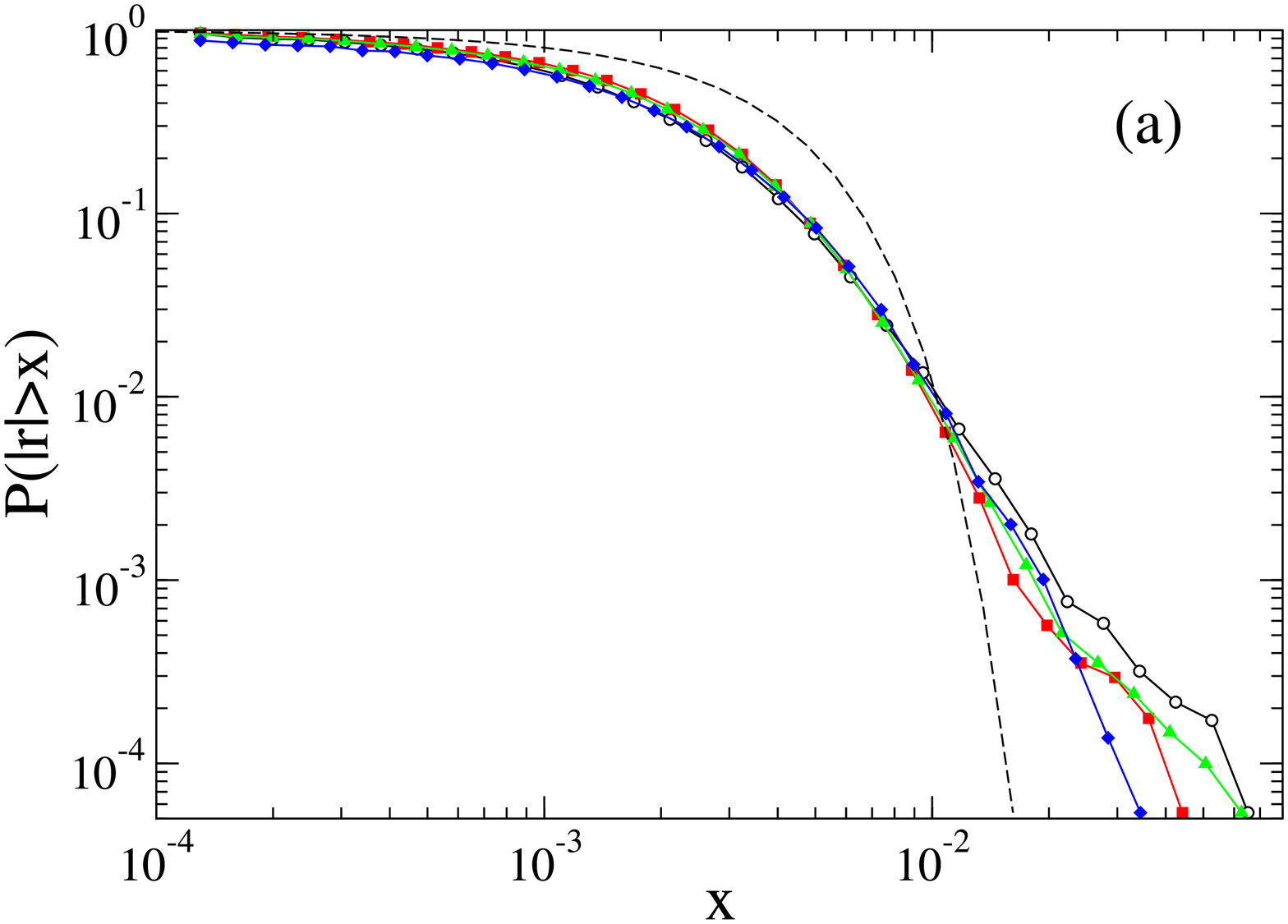}
     \vspace{0.1in}
     
     \includegraphics[height=3.2in, width=1.9in,angle=-90]{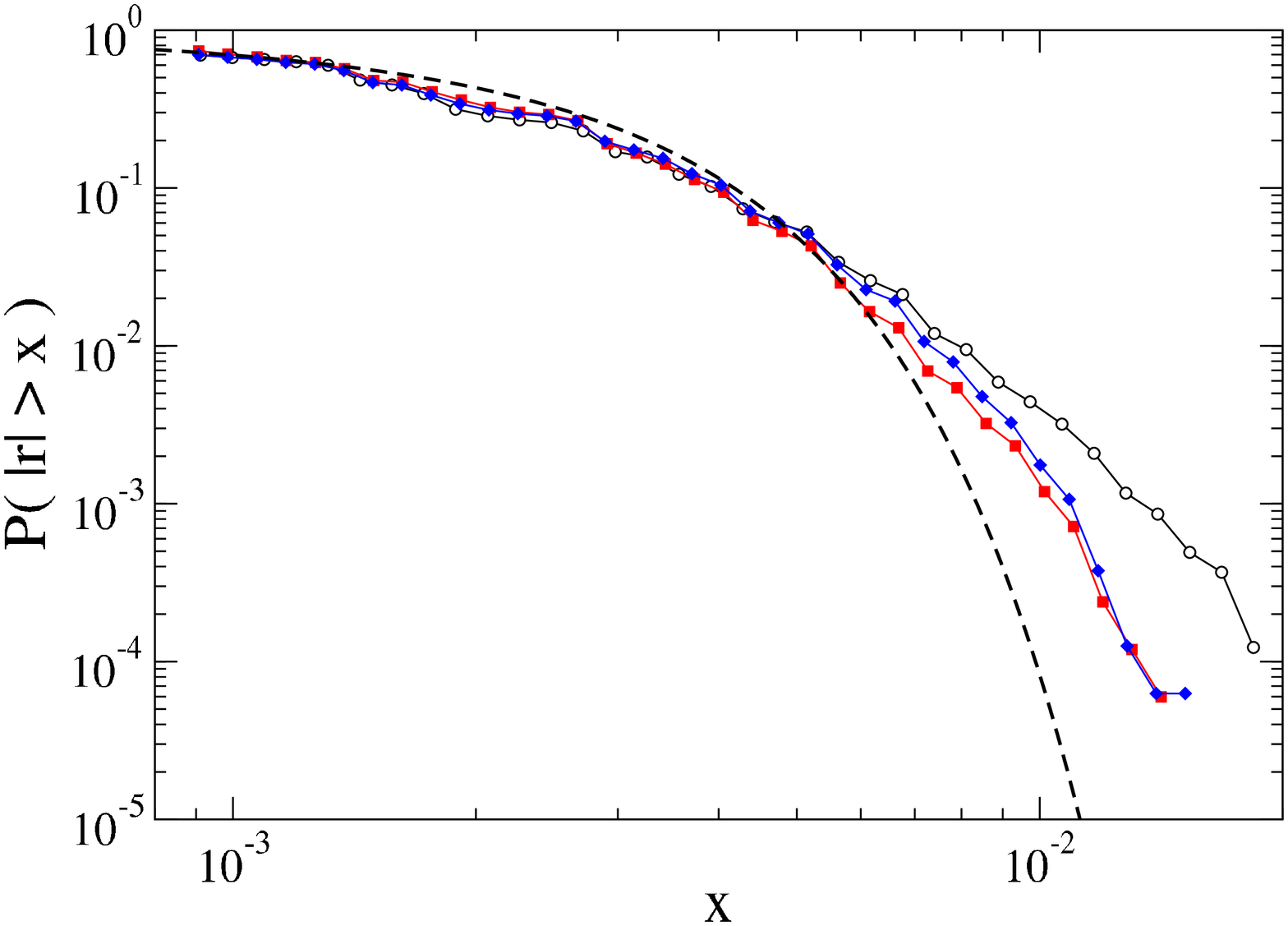}
     \vspace{0.3in}
     
     \includegraphics[height=3.2in, width=1.9in, angle=-90]{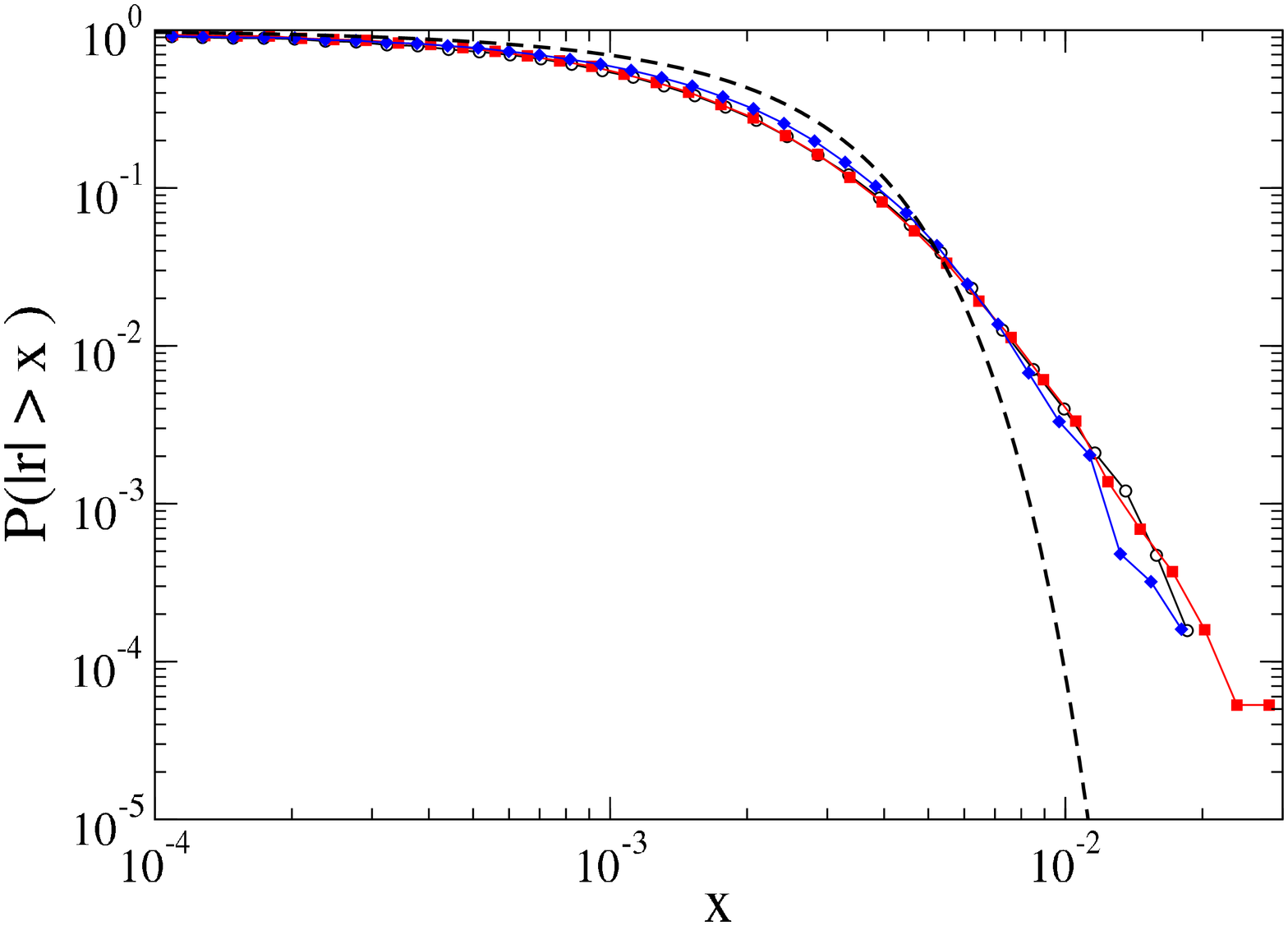}
     \hspace{0.01in} \vspace{0.1in}

     \caption{Cumulative distribution of absolute (log) returns $P(|r| > x)$ under several different time clocks, plotted on double logarithmic scale. (a, top) The LSE stock AZN (b, middle) The NYSE stock Procter \& Gamble (PG) in the 1/8 dollar tick size period. (c, bottom) PG when the tick size is a penny. The time clocks shown are real time (black circles), transaction time (red squares), and volume time (blue diamonds). For comparison we also show a normal distribution (dashed black line) with the same mean and variance as in transaction time. For AZN we also show volume time for the on-book market only (green triangles). The real time returns are $15$ minutes in every case; transaction and volume time intervals are chosen to match this. For both LSE and NYSE2 both the transaction time and volume time distributions are almost identical to real time, even in the tails. For PG in the NYSE1 data set the deviations in the tail from real time are noticeable, but they are still well away from the normal distribution.}
    \label{volatilityEventVsReal}
  \end{center}
\end{figure}

It is difficult to assign error bars to the empirical distributions due to the fact that the data are long-memory (see Section~\ref{longMemory}), so that standard methods tend to seriously underestimate the error bars.  However, it is apparent that for the LSE stock Astrazenca and for the NYSE stock Procter and Gamble in the period where the tick size is a penny, the differences between real time, volume time, and transaction time are minimal.  The situation is more complicated for Procter and Gamble in the period where the tick size is $1/8$.  The distributions for volume and transaction time are nearly identical.  While they are still clearly heavy tailed, the tails are not as heavy as they are in real time.  This leaves open the possibility that during the period of the $1/8$ tick size in the NYSE, volume or transaction frequency fluctuations may have played a role in generating heavy tails in the distribution.  Nonetheless, even here it is clear that they are not the only factor generating heavy tails.

\begin{figure}[ptb]
   \begin{center}

     \hspace{0.01in} \vspace{0.2in}
     
     \includegraphics[height=1.9in, width=3.2in]{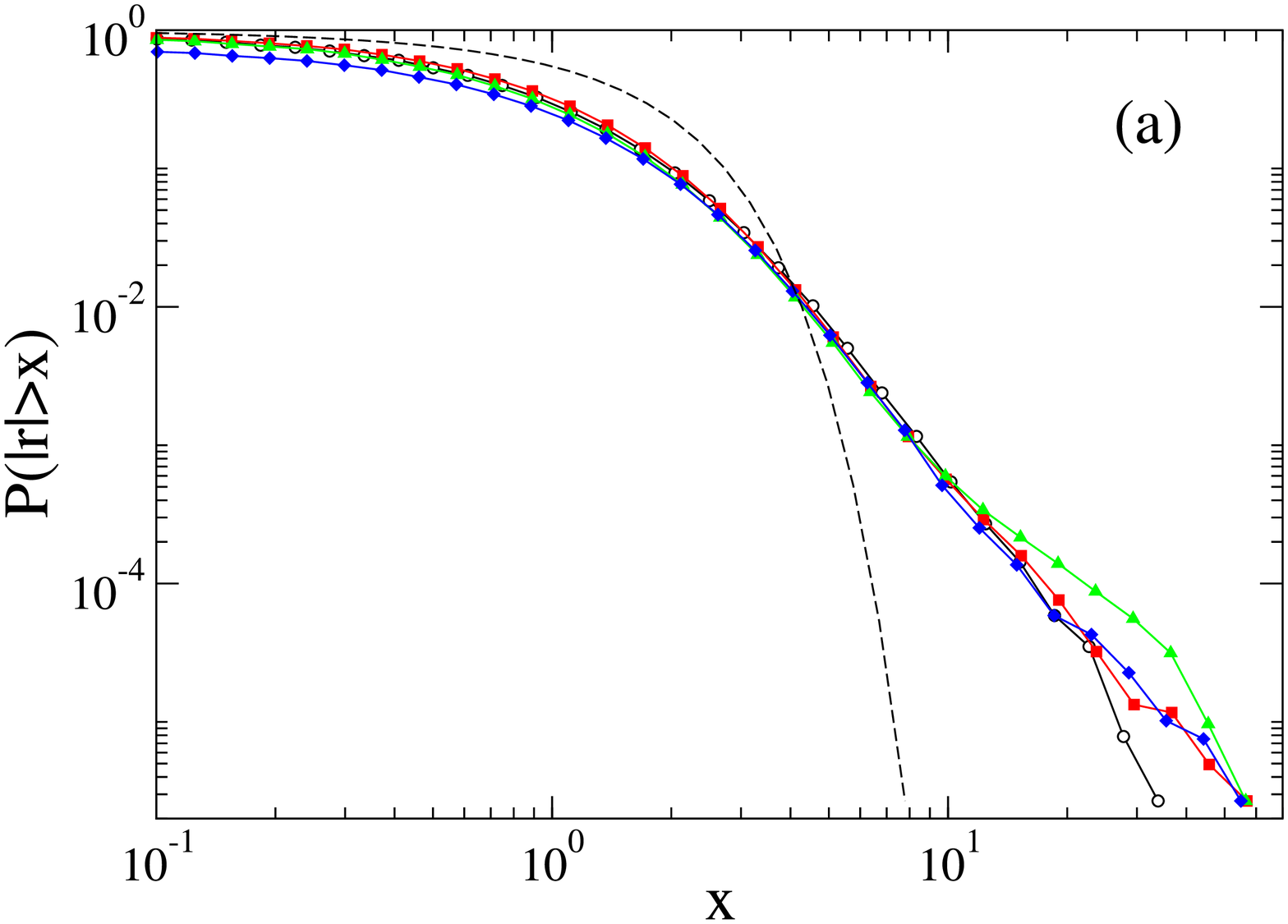}
     
     \hspace{0.01in} \vspace{0.05in}
     
     \includegraphics[height=3.2in, width=1.9in, angle=-90]{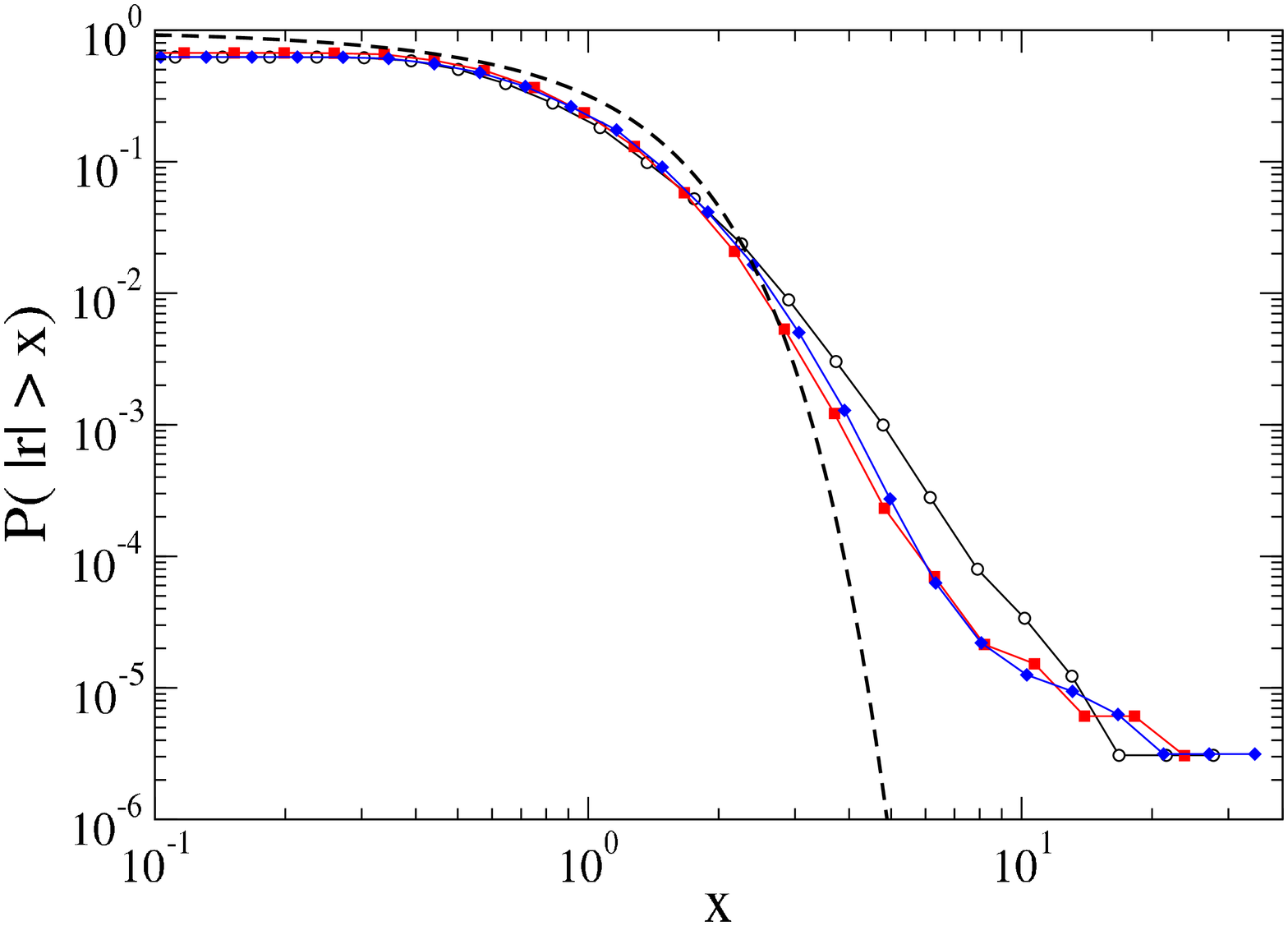}
     
     \hspace{0.01in} \vspace{0.2in}
     
     \includegraphics[height=3.2in, width=1.9in, angle=-90]{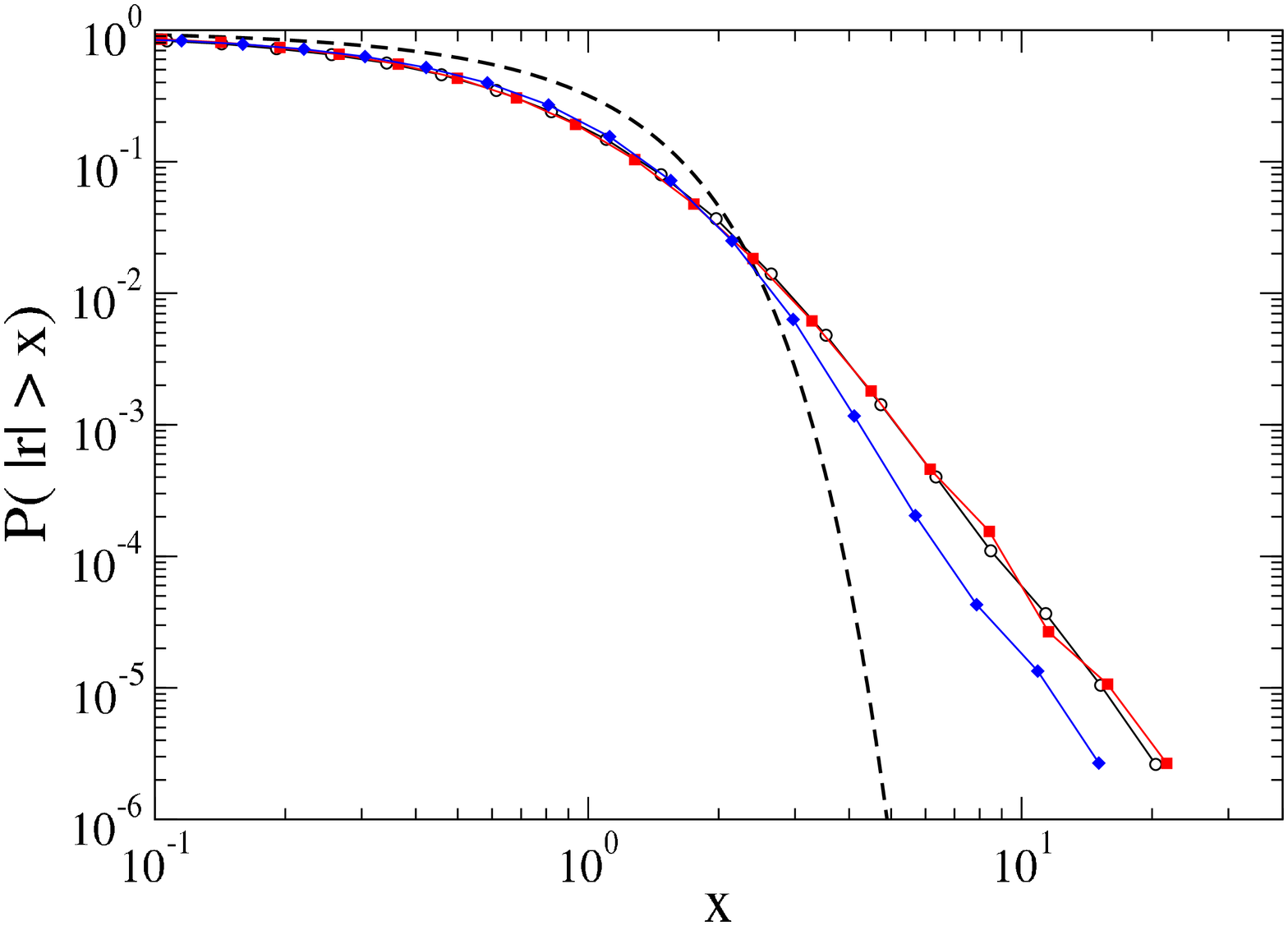}

     \hspace{0.01in} \vspace{0.1in}

     \caption{Cumulative distribution of normalized absolute (log) returns $P(|r| > x)$ averaged over all the stocks in each data set under several different time clocks, plotted on double logarithmic scale. (a, top) LSE, (b, middle) NYSE1, (c, bottom) NYSE2. The time clocks shown are real time (black circles), transaction time (red squares), and volume time (blue diamonds). For the LSE we also show volume time for the on-book market only (green triangles). For comparison we also show a normal distribution (dashed black line) with the same mean and variance as in transaction time.}
    \label{volatilityEventVsReal}
  \end{center}
\end{figure}

To see whether these patterns are true on average, in Figure 5 we show similar plots, except that here we compute an average of the distributions over each of our data sets. We first normalize the data so that the returns for each stock have the same standard deviation, and then we average the distributions for each stock with equal weighting.  For many purposes this can be a dangerous procedure; as we discuss in Section~\ref{GabaixSection}, if different stocks have different tail behaviors this can lead to a biased estimate of the tail exponent.  However, in this case we are just making a qualitative comparison, and we do not think this is a problem.

For the LSE stocks all the distributions appear to be the same except for statistical fluctuations.  The volume time distribution based on on-book volume appears to have somewhat heavier tails than the real time distribution, but for the largest values it joins it again, suggesting that these variations are insignificant.  Throughout part of its range the tails for the volume and transaction time distributions for the NYSE1 data set seem to be less heavy, but for the largest fluctuations they rejoin the real time distribution, again suggesting that the differences are statistically insignificant.  For NYSE2 transaction time is almost indistinguishable from real time; there is a slight steepening of the tail for volume time, but it would be difficult to argue that there is any statistically significant difference in the tail exponents.  These results demonstrate that fluctuations in volume or number of transactions do not have much effect on the return distribution, and are not sufficient to explain its heavy tails.

\subsection{Comparison to work of Ane and Geman}

Our results appear to contradict those of Ane and Geman (2000).  They studied the hypothesis $Y(t) = X(\tau(t))$, where $Y(t)$ is the stochastic process generating real returns $r(t)$, $X$ is a Brownian motion, and $\tau(t)$ is a stochastic time clock.  They developed a method to find the stochastic time clock $\tau$ that would satisfy this assumption, by assuming that the increments $\Delta \tau(t) = \tau(t) - \tau(t-1)$ of the timeclock are IID and computing the moments of its distribution.  They then applied this method to data from two stocks, Cisco and Intel, from January 2 to December 31 in 1997.  Using several different time scales ranging from one to fifteen minutes they computed the moments of $\Delta \tau$ needed to make $X$ a Brownian motion, and compared them to the moments of transaction time increments $\Delta \tau_\theta$ (which is just the number of transactions in a given time). They found that the two sets of moments were very close.  This indicates that the non-normal behavior of the distribution of transactions is sufficient to explain the non-normality of stock returns.

\begin{figure}[ptb]
   \begin{center}
     \includegraphics[scale=0.6,angle=-90]{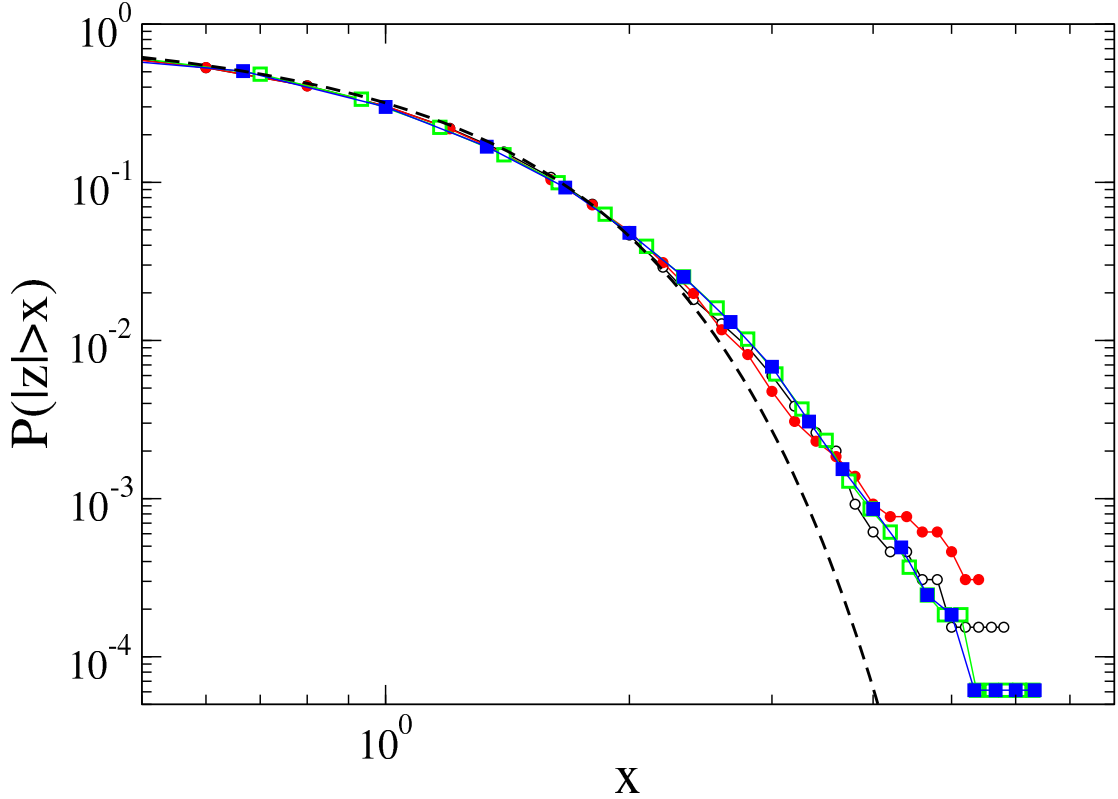}
     \caption{Cumulative distribution of transaction normalized absolute (log) returns $P(|z(t)|)$, where $z(t) = y(t)/\sqrt{N(t)}$, for the stocks Cisco (black circles) and Intel (red filled circles) during the period of the Ane and Geman study, Procter \& Gamble of NYSE1 (green squares), and of NYSE2 (blue filled squares).  These are plotted on double logarithmic scale.  For comparison we also show a normal distribution (dashed black line).  All four distributions are roughly the same, and none of them are normal.}
    \label{cdfComparison}
  \end{center}
\end{figure}

Because this conclusion is quite different from ours, we test their hypothesis on the same data that they used, attempting to reproduce their procedure for demonstrating normality of returns in transaction time\footnote{We are attempting to reproduce Figure~4 of Ane and Geman (2000); it is not completely clear to us what algorithm they used to produce this.  The procedure we describe here is our best interpretation, which also seems to match with that of Deo, Hseih and Hurvich (2005).}.   We construct the returns $r(t ) = p(t) - p(t - \Delta t)$ with $\Delta t = 15$ minutes, and count the corresponding number of transactions $N(t)$ in each interval.   We then create transaction normalized returns $z(t) = r(t)/\sqrt{N(t)}$.  We construct distributions for Intel and Cisco on the same time periods of their study, and for comparison we also do this for Procter and Gamble over the time periods used in our study for the NYSE1 and NYSE2 data sets.  The results are shown in Figure~\ref{cdfComparison}.
In the figure it is clear that all four distributions are quite similar, and that none of them are normal.  As a further check we performed the Bera-Jarque test for normal behavior, getting $p$ values for accepting normality for every stock less than $10^{-23}$.  We do not know why our results differ from those of Ane and Geman, but we note that Li has concluded that their moment expansion contained a mistake (Li 2005). 

\subsection{Implications for the theory of Gabaix et al.\label{GabaixSection}}

Our results here are relevant for the theory for price formation recently proposed by Gabaix et al. (2003).  They have argued that the price impact of a large volume trade is of the form
\begin{equation}
r_i = k \epsilon_i V_i^{1/2} + u_i,
\label{priceImpact}
\end{equation}
where $r_i$ is the return generated by the $i^{th}$ trade, $\epsilon_i$ is its sign ($+1$ if buyer-initiated and $-1$ if seller initiated), $V_i$ is its size, $k$ is a proportionality constant and $u_i$ is an uncorrelated noise term.  Under the assumption that all the variables are uncorrelated, if the returns are aggregated over any given time period the squared return is of the form 
\begin{equation}
E[r^2 | V] = k^2V + E[u^2], 
\label{r2}
\end{equation}
where $r = \sum_i^N r_i$, $V = \sum_i^N V_i$, $E[u^2] = \sum_i^N E[u_i^2]$, and $N$ is the number of transactions, which can vary.   They have hypothesized that equation~\ref{r2} can be used to infer the tail behavior of returns from the distribution of volume.   Their earlier empirical work found that the distribution of volume has heavy tails that are asymptotically of the form $P(V > v) \sim v^{-\alpha}$, with $\alpha \approx 1.5$ (Gopikrishnan et al. 2000).  These two relations imply that the tails of returns should scale as $P(r > R) \sim R^{-2 \alpha}$.  

This theory has been criticized on several grounds.  The most important points that have been raised are that Equation~\ref{priceImpact} is not well supported empirically, that $\epsilon_i$ are strongly positively correlated with long-memory so that the step from Equation~\ref{priceImpact} to Equation~\ref{r2} is not valid, and that at the level of individual transactions $P(r_i > x | V_i)$ only depends very weakly on $V_i$ (Farmer and Lillo 2004, Farmer et al. 2004 -- see also the rebuttal by Plerou et al. 2004).  Our results here provide additional evidence\footnote{In Farmer et al. (2004) we showed that the returns were independent of volume at the level of individual transactions, and only for the on-book market of the LSE.  Here we show this for 15 minute intervals, we control for volume in both the on-book and off-book markets, and we also study NYSE stocks.}.  The distribution of returns in volume time plotted in Figures~4 and 5 is just the distribution of returns when the volume is held constant, and can be written as $P(r > x | V = \Delta V )$, where $\Delta V$ is the sampling interval in volume time as described in Section~\ref{volumeTime}.   We can express their hypothesis in a more general different form by squaring Equation~\ref{priceImpact} and rewriting Equation~\ref{r2} in the form $r^2 = k^2 V + w$, where $w \equiv r^2 - k^2 V$ is a noise term that captures all the residuals; this allows for the possibility that $w \neq u^2$.   When $V$ is held constant $r$ should be of the form $r = \sqrt{C + w}$, where $C = k^2V$ is a constant.  According to their theory the noise term $w$ should be unimportant for the tails.  Instead we see that in every case $P(r > x| V)$ is heavy tailed, and in many cases it is almost indistinguishable from $P(r > x)$.   This indicates that $w$ is not negligible, but is rather the dominant factor.

\begin{figure}[ptb]
   \begin{center}

     \includegraphics[scale=0.4,angle=-90]{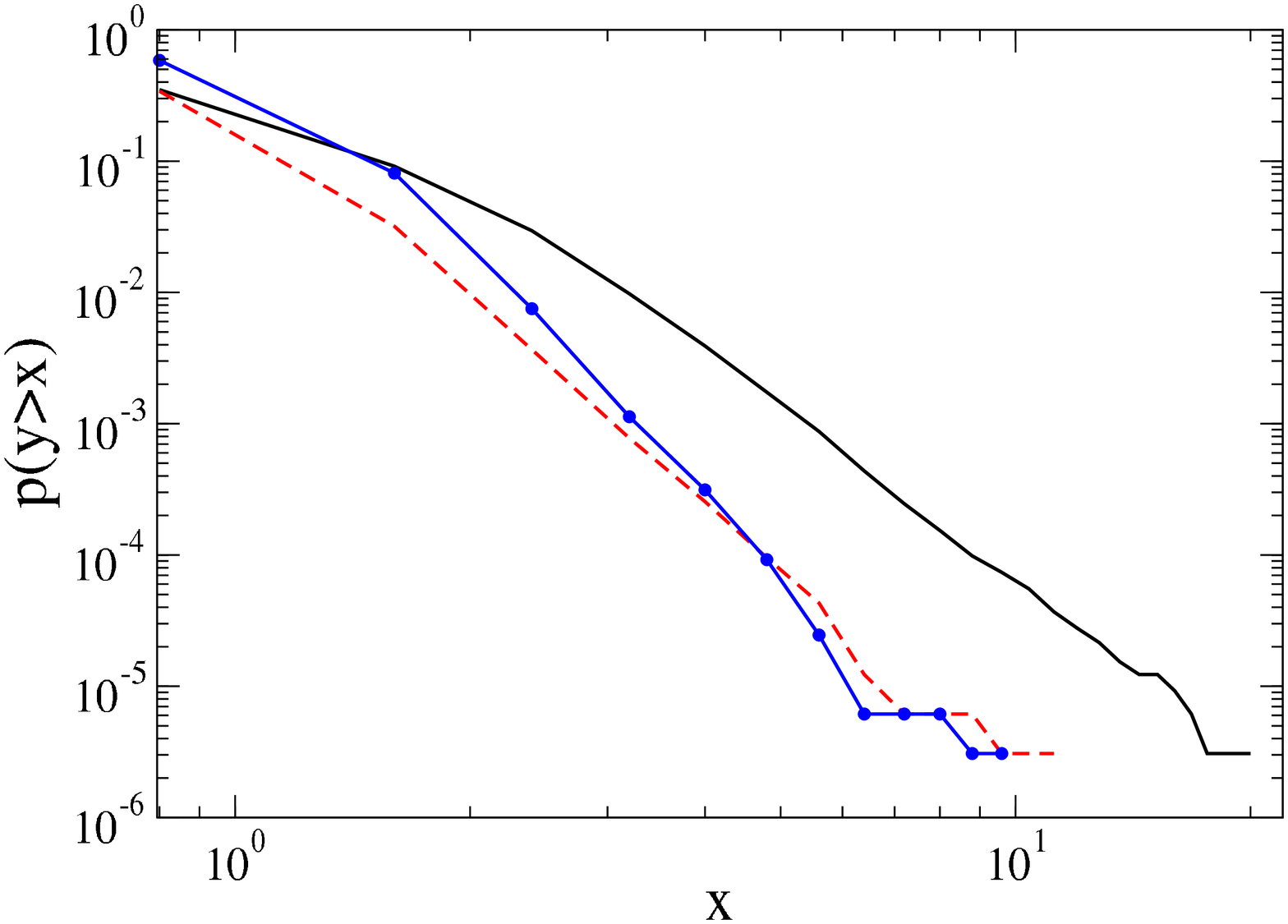}
     \hspace{0.01in} \vspace{0.4in}

     \includegraphics[scale=0.4,angle=-90]{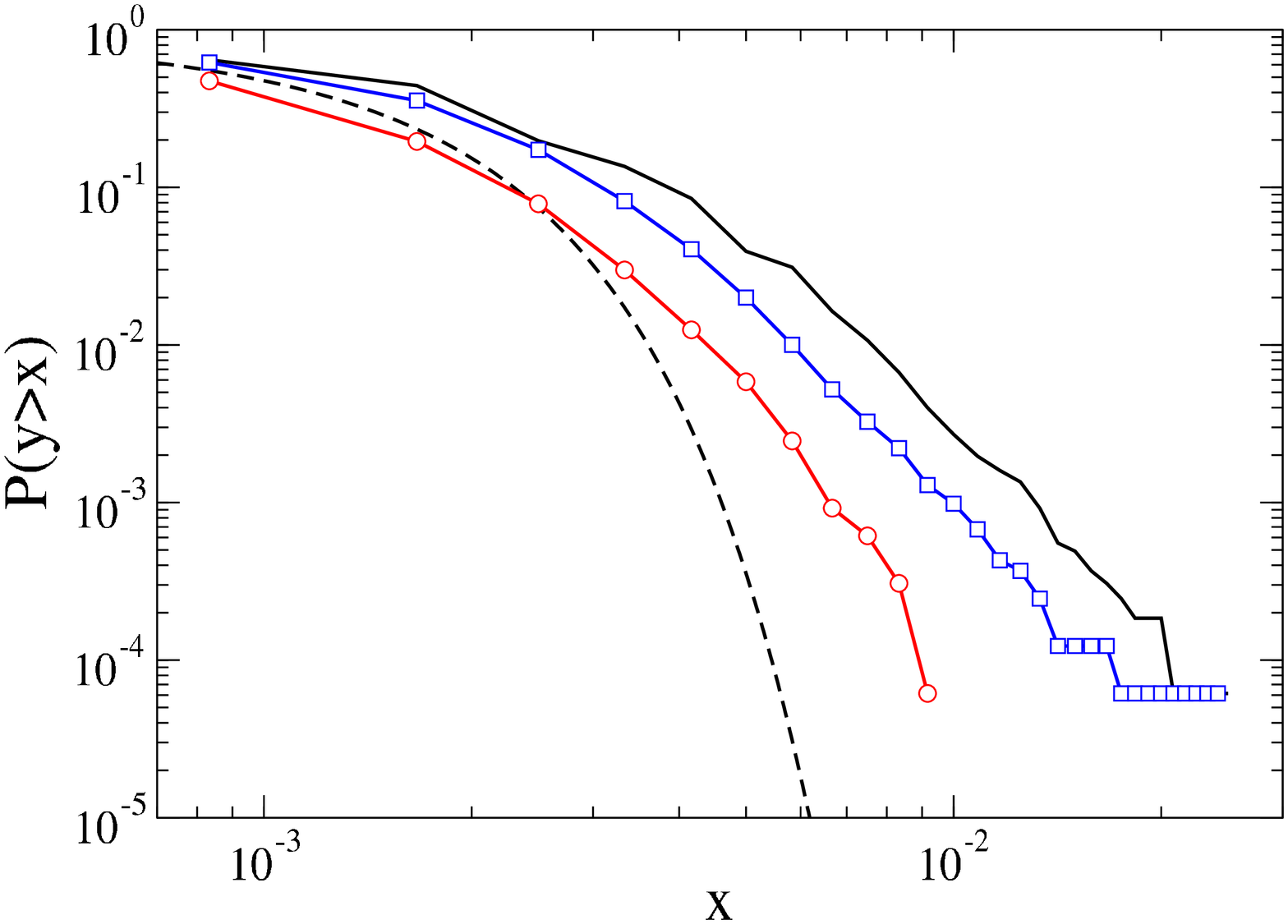}

     \caption{Test of the theory of Gabaix et al.  In (a) we compare the empirical return distribution to the prediction of the theory of Gabaix et al. for the NYSE1 data set.  The solid black curve is the empirical distribution $P(r > x)$, the dashed red curve is the distribution of predictions $\hat{r}$ under the assumption that $\beta = 1.2$, and the blue curve with circles is same thing with $\beta = 1$.  In (b) we compare the empirical distribution $P(r > x)$ (black), the distribution of residuals $P(\eta > x)$ (blue squares), and the predicted return under the theory of Gabaix et al., $P(\hat{r} > x)$ (red circles).  The dashed black line is a normal distribution included for comparison.}
     \label{Gabaix}
     \end{center}
     \end{figure}

This is puzzling because Gabaix et al. have presented empirical evidence that when a sufficient number of stocks are aggregated together $E[r^2 | V]$ satisfies Equation~\ref{r2} reasonably well.  To understand how this can coexist with our results, we make some alternative tests of their theory.  We begin by studying $E[r^2 | V]$.  Following the procedure they outline in their paper, using $15$ minute intervals we normalize $r$ for each stock by subtracting the mean and dividing by the standard deviation, and normalize $V$ for each stock by dividing by the mean.  We allow a  generalization of the assumption by letting $E[r^2 | V] = k^2 V^\beta$, i.e. we do not require that $\beta = 1$.  Using least squares regression, for large values of $r$ and $V$ we find $k \approx 0.61$ and $\beta \approx 1.2$. Their theory predicts that $P(r > x) \sim P(kV^{\beta/2} > x)$ for large $x$.  In Figure~\ref{Gabaix}a we use this together with the empirical distribution of volume $P(V)$ to estimate $P(r > x)$ and compare it to the empirical distribution.
Regardless of whether we use $\beta = 1.2$ or $\beta = 1$, the result has a much thinner tail than the actual return distribution\footnote{One reason that we observe a thinner tail is because for this set of stocks we observe thinner tails in volume.  We find $P(V > v) \sim v^{-\alpha}$, with $\alpha$ in the range $2 - 2.4$, clearly larger than the value $\alpha = 1.5$ they have reported. This difference appears to be caused by the portion of the data that we are using.  We remove all trades that occur outside of trading hours, and also remove the first and last trades of the day.  In fact these trades are consistently extremely large, and when we include them in the data set we get a tail exponent much closer to $\alpha = 1.5$.  Since we are using volume and returns consistently drawn from matching periods of time, for the purpose of predicting returns from volume under their theory this should not matter.  Note that for the returns we observe tails roughly the same as they have reported; for the LSE they are even heavier, with most stocks having $\alpha$ significantly less than three.}.
 
In Figure~\ref{Gabaix}b we perform an alternative test.   We estimate the predicted return in each 15 minute interval according to their theory as $\hat{r} = A + kz$, where $z = \sum_i \epsilon_i V_i^{1/2}$, using least squares approximation to estimate $A$ and $k$ in each interval. This can be used to compute a residual $\eta = r - \hat{r}$.  We then compare the distributions of $r$, $\hat{r}$, and $\eta$; according to their theory $\hat{r}$ should match the tail behavior and $P(\eta)$ should be unimportant.  Instead we find the opposite.  $P(\eta)$ is closer to the empirical distribution $P(r)$, roughly matching its tail behavior, while $P(\hat{r})$ falls off more rapidly -- for the largest fluctuations it is about an order of magnitude smaller than the empirical distribution.  This indicates that the reason their theory fails is that the heavy-tailed behavior it predicts is dominated by the even heavier-tailed behavior of the residuals $\eta$.  In our earlier study of the LSE we were able to show explicitly why this is true.   There we demonstrated that heavy tails in returns at the level of individual transactions are primarily caused by the distributions of gaps (regions without any orders) in the limit order book, whose size is unrelated to trading volume (Farmer et al. 2004). 

\section{Long-memory \label{longMemory}}

There is mounting evidence that volatility is a long-memory process (Ding, Granger, and Engle 1993, Breidt, Crato and de Lima 1993, Harvey 1993, Andersen et al. 2001).  A long-memory process is a random process with an autocorrelation function that is not integrable. This typically happens when the autocorrelation function decays asymptotically as a power law of the form $\tau^{-\gamma}$ with $\gamma<1$. The existence of long-memory is important because it implies that values from the distant past have a significant effect on the present and that the stochastic process lacks a typical time scale.  A stochastic process that is build out of a sum whose increments have long-memory has anomalous diffusion, i.e. the variance grows as $\tau^{2H}$, where $H > 0.5$, and $H$ is the Hurst exponent, which is defined below.  Statistical averages of long-memory processes converge slowly.

Plerou et al. (2000) demonstrated that the number of transactions in NYSE stocks is a long-memory process.   This led them to conjecture that fluctuations in the number of transactions are the proximate cause of long-memory in volatility.   This hypothesis makes sense in light of the fact that either long-memory or sufficiently fat tails in fluctuations in transaction frequencies are sufficient to create long-memory in volatility (Deo, Hsieh, and Hurvich 2005).  However, our results so far in this paper suggests caution -- just because it is possible to create long-memory with this mechanism does not mean it is the dominant mechanism -- there may be multiple sources of long-memory.  If there are two long memory processes with autocorrelation exponents $\gamma_1$ and $\gamma_2$, if $\gamma_1 < \gamma_2$ the long-memory of process one will dominate that of process two, since for large values of $\tau$ $C_1(\tau) \gg C_2(\tau)$.  We will see that this is the case here -- the long-memory caused by fluctuations in volume and number of transactions is dominated by long-memory caused by other factors.

We investigate the long-memory of volatility by computing the Hurst exponent $H$.  For a random process whose second moment is finite and whose autocorrelation function asymptotically decays as a power law $\tau^{-\gamma}$ with $0<\gamma<1$, the Hurst exponent is $H = 1 - \gamma/2$.  A process has long memory if $1/2 < H < 1$.  We compute the Hurst exponent rather than working directly with the autocorrelation function because it is better behaved statistically.  We compute the Hurst exponent using the DFA method (Peng et al. 1994).  The time series $\{x(t)\}$ is first integrated to give $X(t) = \sum_{i=1}^t x(i)$. For a data set with $n$ points,  $X(t)$ is then broken into $K$ disjoint sets $k = 1, \ldots, K$ of size $L \approx n/K$.  A polynomial $Y_{k,L}$ of order $m$ is then fit for each set $k$ using a least squares regression.  For a given value of $L$ let $D(L)$ be the root mean square deviation of the data from the polynomials averaged over the $K$ sets, i.e. $D(L) = {(1/K \sum_{k,i}  (X(i) - Y_{k,L}(x(i))^2)}^{1/2}$.  The Hurst exponent is formally defined as $H = \lim_{L \to \infty} \lim_{n \to \infty} \log D(L)/\log L$.  In practice, with a data set of finite $n$ the Hurst exponent is measured by regressing $\log D(L)$ against $\log L$ over a region with $L_{min} < L <  n/4$.   $H$ is the slope of the regression.

\begin{figure}[ptb]
   \begin{center}
     \includegraphics[scale=0.4]{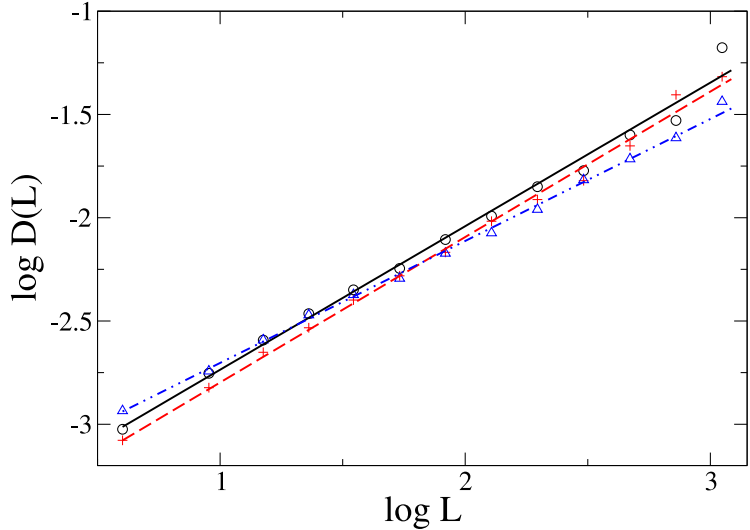}
     \caption{Computation of Hurst exponent of volatility for the LSE stock Astrazeneca.  The logarithm of the average variance $D(L)$ is plotted against the scale $\log L$.  The Hurst exponent is the slope.  This is done for real time volatility $\nu$ (black circles), transaction time volatility $\nu_\theta$ (red crosses), and shuffled transaction real time volatility $\tilde{\nu}_\theta$ (blue triangles).  The slopes in real time and transaction time are essentially the same, but the slope for shuffled transaction real time is lower, implying that transaction fluctuations are not the dominant cause of long-memory in volatility.}
    \label{hurstScaling}
  \end{center}
\end{figure}

To test the hypothesis that the number of transactions drives the long-memory of volatility we compute the Hurst exponent of the real time volatility,  $H(\nu)$, and compare it to the Hurst exponent of the volatility in transaction time, $H(\nu_\theta)$, and the Hurst exponent in shuffled transaction real time, $H(\tilde{\nu}_\theta)$.  If the long-memory of transaction time is the dominant cause of volatility then we should see that $H(\nu) \approx H(\tilde{\nu}_\theta)$, due to the fact that $\tilde{\nu}$ preserves the number of transactions in each $15$ minute interval, and we should also see that $H(\nu) > H(\nu_\theta)$, due to the fact that $\nu_\theta$ holds the number of transactions constant in each interval, so it should not display as much clustered volatility.   We illustrate the scaling diagrams used to compute these three Hurst exponents for the LSE stock Astrazeneca in Figure~\ref{hurstScaling}.
This figure does not support the conclusion that transaction time fluctuations are the proximate cause of volatility.  We find $H(\nu) \approx 0.70 \pm 0.07$, $H(\nu_\theta) \approx 0.70 \pm 0.07$, and $H(\tilde{\nu}_\theta) \approx 0.59 \pm 0.03$.  Thus for Astrazenca it seems the opposite is true, i.e. $H(\nu) \approx H(\nu_\theta)$ and $H(\nu) > H(\tilde{\nu}_\theta)$.  While it is true that $H(\tilde{\nu}_\theta) > 0.5$, which means that fluctuations in transaction frequency contribute to long-memory, the fact that $H(\nu_\theta) > H(\tilde{\nu}_\theta)$ means that this is dominated by even stronger long-memory effects that are caused by other factors.

Note that the quoted error bars for $H$ as based on the assumption that the data are normal and IID, and are thus much too optimistic.   For a long-memory process such as this the relative error scales as $n^{(H-1)}$ rather than $n^{-1/2}$, and estimating accurate error bars is difficult.  The only known procedure is the variance plot method (Beran, 1992), which is not very reliable and is tedious to implement.  This drives us to make a cross-sectional test, where the consistency of the results and their dependence on other factors will make the statistical significance quite clear.

\begin{figure}[ptb]
   \begin{center}
     \includegraphics[scale=0.6,angle=-90]{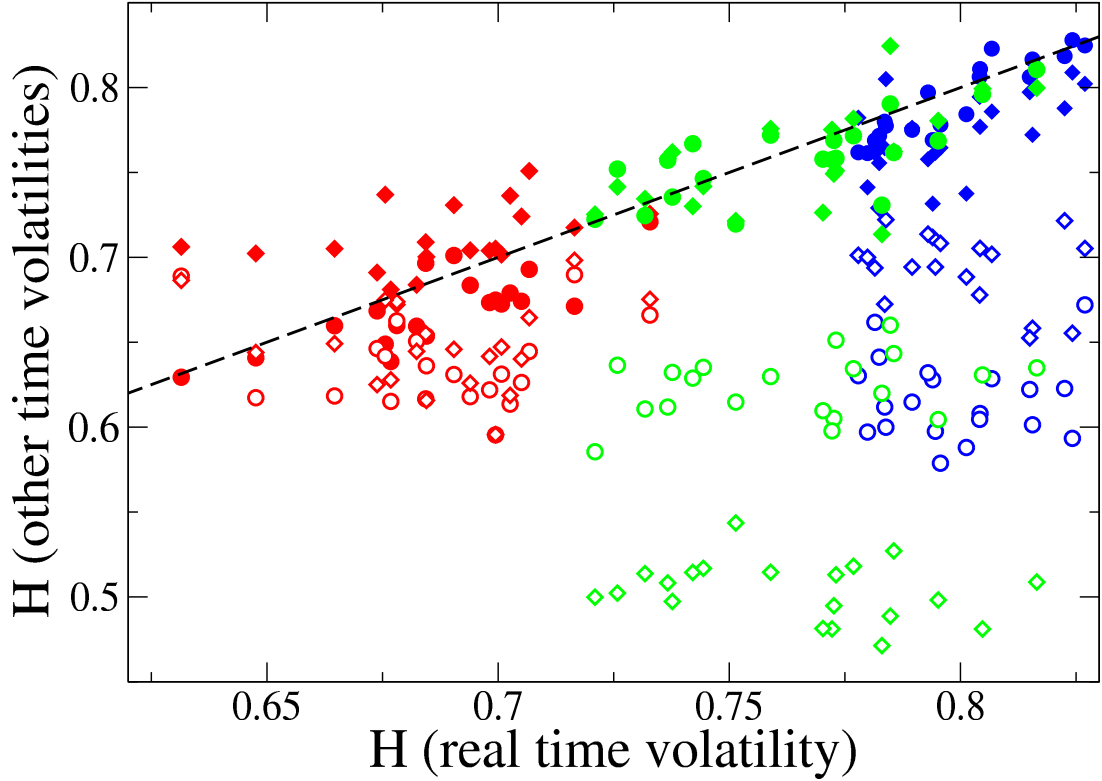}
     \caption{Hurst exponents for alternative volatilities vs. real time volatility.  Each point corresponds to a particular stock and alternative volatility measure.  NYSE1 stocks are in red, LSE in green, and NYSE2 in blue.    The solid circles are transaction time, open circles are shuffled transaction real time, solid diamonds are volume time, and open diamonds are shuffled volume real time.  The black dashed line is the identity line. The fact that transaction time and volume time Hurst exponents cluster along the identity line, whereas almost all of the shuffled real time values are well away from it, shows that neither volume nor transaction fluctuations are dominant causes of long-memory.}
    \label{hurstScatter}
  \end{center}
\end{figure}

To test the consistency of our results above we compute Hurst exponents for all the stocks in each of our three data sets.  In addition, to test whether fluctuations in volume are important, we also compute the Hurst exponents of volatility in volume time, $H(\nu_v)$ and in shuffled volume real time, $H(\tilde{\nu}_v)$.  The results are shown in Figure~\ref{hurstScatter}, where we plot the Hurst exponents for $H(\nu_\theta)$,  $H(\tilde{\nu}_\theta)$, $H(\nu_v)$, and $H(\tilde{\nu}_v)$ against the real time volatility $H(\nu)$ for each stock in each data set.
Whereas the Hurst exponents in volume and transaction time cluster along the identity line, the Hurst exponents for shuffled real time are further away from the identity line and are consistently lower in value.  This is seen at a glance in the figure by the fact that the solid marks are clustered along the identity line whereas the open marks are scattered away from it.  The results are quite consistent -- out of the $60$ cases shown in Figure~\ref{hurstScatter}, only one has $H(\tilde{\nu}_v) > H(\nu_v)$ and only one has $H(\tilde{\nu}_\theta) > H(\nu_\theta)$.

As a further test we perform a regression of the form $H_i^{(a)} = a + bH_i^{(r)}$, where $H_i^{(a)}$ is an alternative Hurst exponent and $H_i^{(r)}$ is the Hurst exponent for real time volatility for the $i^{th}$ stock.  We do this for each data set and each alternative volatility measure.   The results are presented in Table~\ref{hurstTable}.  For volume and transaction time all of the slope coefficients $b$ are positive, all but one at the $95\%$ confidence level.  This shows that as the long-memory of real time volatility increases the volatility when the volume or number of transactions is held constant increases with it.   In contrast for the shuffled volume real time measures four out of six cases have negative slopes and none of them are statistically significant.  This shows that the long-memory of real time volatility is not driven by the volume or the transaction frequency.  The order of events is important -- more important than their number.  The $R^2$ values of the regressions are all higher for volume or transaction time than for their corresponding shuffled values, and the mean distances from the identity line are substantially higher.  These facts taken together make it clear that the persistence of real time volatility is driven more strongly by other factors than by volume or transaction fluctuations.
\begin{table}
%\begin{center}
\begin{tabular}{l|l|cccc}
volatility&market&$a$&$b$&$d~(\times 100)$&$R^2~(\times 100)$\\
\hline
&NYSE1&$~~0.15\pm 0.10$&$~~0.76\pm 0.20$&$1.4$&$61$\\
$\nu_\theta$&NYSE2&$-0.27\pm0.11$&$~~1.34\pm0.14$&$0.76$&$82$\\
&LSE&$~~0.26\pm0.11$&$~~0.65\pm0.15$&$1.1$&$52$\\
\hline
&NYSE1&$~~0.46\pm 0.12$&$~~0.35\pm 0.19$&$1.4$&$16$\\
$\nu_v$&NYSE2&$~~0.17\pm0.23$&$~~0.75\pm0.29$&$2.0$&$27$\\
&LSE&$~~0.25\pm0.16$&$~~0.67\pm0.21$&$1.6$&$37$\\
\hline
&NYSE1&$~~0.65\pm 0.18$&$-0.03\pm 0.25$&$4.0$&$8.3~10^{-2}$\\
$\tilde\nu_\theta$&NYSE2&$~~0.59\pm0.29$&$~~0.03\pm0.35$&$13$&$4.6~10^{-2}$\\
&LSE&$~~0.47\pm0.12$&$~~0.19\pm0.16$&$10$&$8.1$\\
\hline
&NYSE1&$~~0.67\pm 0.18$&$-0.04\pm 0.26$&$3.2$&$1.2~10^{-1}$\\
$\tilde\nu_v$&NYSE2&$~~1.11\pm0.25$&$-0.52\pm0.31$&$7.3$&$14$\\
&LSE&$~~0.64\pm0.11$&$-0.18\pm0.15$&$18$&$7.5$\\
\hline
 \end{tabular}
% \end{center}
\caption{A summary of results comparing Hurst exponents for alternative volatilities to real time volatility.  
We perform regressions on the results in Figure~\ref{hurstScatter} of the form $H_i^{(a)} = a + bH_i^{(r)}$, where $H_i^{(a)}$ is an alternative Hurst exponent and $H_i^{(r)}$ is the Hurst exponent for real time volatility for the $i^{th}$ stock.  We do this for each data set and each of the exponents $H(\nu_\theta)$ (transaction time),  $H(\tilde{\nu}_\theta)$ (shuffled transaction real time), $H(\nu_v)$ (volume time), and $H(\tilde{\nu}_v)$ (shuffled volume real time).  $d$ is the average distance to the identity line $H_i^{(a)}  = H_i^{(r)}$ and $R^2$ is the goodness of fit of the regression.  We see that in the top two rows $b$ is positive and statistically significant in all but one case, in contrast to the bottom two rows.  This and the fact that $d$ is much smaller in the top two rows makes it clear that neither volume nor transactions are important causes of long-memory.}
\label{hurstTable}
\end{table}

It is interesting that the data set appears to be the most significant factor determining $H$.  For the NYSE the real time Hurst exponents during the $1/8$ tick size period are all in the range $0.62 < H < 0.73$, whereas in the penny tick size period they are in the range $0.77 < H < 0.83$.   Thus the Hurst exponents for the two periods are completely disjoint -- all the values of $H$ during the penny tick size period are higher than any of the values during the $1/8$ tick size period.  The LSE Hurst exponents are roughly in the middle, spanning the range $0.72 < H < 0.82$.  It is beyond the scope of the paper to determine why this is true, but this suggests that changes in tick size or other aspects of market structure are very important in determining the strength of the persistence of volatility.

\section{Conclusions}

The idea that clustered volatility and heavy tails in price returns can be explained by fluctuations in transactions or volume is seductive in its simplicity.  Both transaction frequency and volume are strongly positively correlated with volatility, and it is clear that fluctuations in volume (or transaction frequency) can cause both clustered volatility and heavy tails, so this might seem to be an open and shut case.  Our main result in this paper is to show that this is not true, at least for the data that we have studied here.  For these data the effects of volume and transaction frequency are dominated by other effects.  We have shown this for three different properties, the contemporaneous relationship with the size of price changes, the long-memory properties of volatility, and the distribution of returns.  The results have been verified with three different large data sets, with tens of millions of transactions, and changes in both the time period and the market.  The story is quite consistent in each case, with only a few minor variations.

This can be viewed as a tale of competing effects.  We do not dispute that volume and transaction frequency affect prices, but for these data other effects are more important.  If there were no other competing effects, fluctuations in volume and transaction would indeed cause clustered volatility and heavy tails.   However in this case the clustered volatility would not be as strong, it would not be as persistent, and there would be fewer extreme price returns.  It is useful to think about this in the context of competing power laws.  For the long-memory properties of volatility, for example, we have shown that volume and transaction frequency effects by themselves give rise to long-memory with Hurst exponents $H(\tilde{\nu}_v)$ and $H(\tilde{\nu}_\theta)$ that are smaller than those of realtime volatility.  This means that volume and transaction frequency fluctuations make contributions to the autocorrelation of volatility that are asymptotically of the form $\tau^{-\gamma_v}$ and $\tau^{-\gamma_\theta}$, where $\gamma_v = 2(1 - H(\tilde{\nu}_v))$ and $\gamma_\theta =  2(1 - H(\tilde{\nu}_\theta))$.  Let $\gamma_r$ be the exponent characterizing the autocorrelation of real time volatility, i.e. $\gamma_r = 2(1 - H(\nu))$.  We have shown that $\gamma_r < \gamma_v$ and also that $\gamma_r < \gamma_\theta$.  This implies that for long times $\tau^{-\gamma_r} \gg \tau^{-\gamma_v}$ and $\tau^{-\gamma_r} \gg \tau^{-\gamma_\theta}$, i.e. that the persistence of volatility is dominated by long-memory effects other than those induced by volume and transaction frequency.  We have also shown that fluctuations in volume and transaction frequency give rise to tails that drop off more steeply than those of real price returns, and thus for large values make a negligible contribution to the heavy tails.   Our conclusion is while volume and transaction frequency have a role to play in price formation, for these data sets they are supporting actors rather than lead players.  It is of course possible that there are other data sets where these roles are reversed, but this is not what we see here; for example in our study of long-memory, out of the 60 cases we examined, we only observed one in which $H(\tilde{\nu}_v) > H(\nu_v)$, and one in which $H(\tilde{\nu}_\theta) > H(\nu_\theta)$.  

One of the intriguing results that is hinted at but not fully explored here is a possible effect of market structure.  We see indications that transaction frequency plays a larger role in the NYSE1 data set, when the tick size was $1/8$, than in the NYSE2 data set, when the tick size was a penny.   When returns are measured in transaction time rather than real time the distribution of price returns is almost unaffected in the NYSE2 data set; this is not so clear in the NYSE1 data set where there is some indication that the tail behavior is not as strong.  This conclusion is not firm due to the difficulty of ensuring statistically significant behavior in the tails in the presence of long-memory, but the difference is suggestive.  There are also other indications of shifts in behavior between the two data sets:  The results in Figure~3 suggest that volume plays a more important role in the NYSE2 data set than in the NYSE1, whereas the situation is reversed for transaction frequency.  The enhanced role for volume is consistent with the results on return distributions in Figures~4 and 5.  The most striking result along these lines, which is clearly statistically significant, is that the long-memory of volatility is more persistent in the NYSE2 data set than in the NYSE1 data set, which is evident in Figure~\ref{hurstScatter}.  It is not clear at this stage whether these changes are due to changes in tick size, whether they are due to some other aspect of market structure, or whether they could be caused by changes in investor behavior, but the coincidence with tick size is suggestive and deserves further investigation.

The most interesting aspect of our results is their implication for price formation.  If  volume and transaction frequency are not the primary proximate cause of price movements, then what is?  It is surprising that after so many years of empirical work the answer to this question remains unknown.  The work we have done so far supports the main themes of the work in Farmer et al. (2004).  There we showed that for the LSE the heavy tails in prices are already present at the microscopic level of individual price fluctuations.   The tails of the distribution are essentially unaltered in transaction time, and the size of individual transactions has only a small effect on the size of price changes.  Taking advantage of the precision of the LSE data set, we looked at midprice changes caused by limit orders placed inside the spread, market orders that remove all the depth at the best prices, and cancellations of the last order at the best price.  We found that they all have essentially the same distribution.  We attribute this fact to the dynamics of gaps in the limit order book, where a gap is defined as a region with no orders.   As price changes are aggregated on longer time scales, the heavy-tailed behavior on long time scales mimics that on shorter time scales, even if the crossover from the body to heavy-tailed behavior occurs at successively larger values.  These results indicate that the main driver of heavy tails is the heavy tailed distribution of individual price changes themselves.  This is driven by order placement and cancellation, and not by volume or transaction frequency.  This is by no means an ultimate cause, but it does provide insight into where such a cause might lie.

This story for the origin of heavy tails is consistent with our work so far studying clustered volatility.  We have shown here that the ordering of transactions is very important in determining clustered volatility.  If we scramble the order of transactions but preserve their frequency, clustered volatility is greatly diminished.  A similar statement applies to volume, making it clear that the size of transactions is also not playing an important role.  What does play an important role is the size of (non-zero) individual price changes and the frequency with which they occur.  We will not present our results here, but our preliminary work suggests that the dominant drivers of both heavy tails and clustered volatility are due to changes in the frequency with which transactions generate non-zero price returns, as well as variations in the size of price returns\footnote{Plerou et al. (\citeyear{Plerou00}) suggested that the properties of individual transactions are the proximate cause of heavy tails (but not clustered volatility).  They seem to have reversed this opinion in (Gabaix et al. \citeyear{Gabaix03}).}.  Both of these effects are strongly correlated in time.  These variables fluctuate because of changes in the balance between liquidity taking and liquidity provision, a theme we have also stressed elsewhere (Farmer et al. 2004, Lillo and Farmer 2004).  This conclusions is reinforced by a new empirical model of order placement and cancellation (Mike and Farmer, 2005).  This model generates return and spread distributions that match those of real data quite well, both in the body and in the tail.  This model suggests that the long-memory of supply and demand (Bouchaud et al. 2004, Lillo and Farmer 2004) is an important effect driving the heavy tails of price returns.

Although based on the results presented here the following remarks are still somewhat speculative, when our results here are combined with those mentioned above, they suggest two basic conclusions about the dominant influences on price formation.  First, the balance between liquidity taking and liquidity provisions is key.  As a result of this imbalance shifting, for a given transaction size prices can change more or less easily, and the size of the changes can vary.  This is typically more important than the number of transactions or their size.  Second, the market displays interesting dynamics on long time scales that cause this imbalance to vary in a highly correlated manner.  We believe that this dynamic is the most important factor driving both clustered volatility and heavy tails.

\acknowledgements{We would like to thank the James S. McDonnell Foundation for their Studying Complex Systems Research Award, Credit Suisse First Boston, Bob Maxfield, and Bill Miller for supporting this research. FL acknowledges support from the research project MIUR 449/97 ``High frequency dynamics in financial markets", the research project MIUR-FIRB RBNE01CW3M ``Cellular Self-Organizing nets and chaotic nonlinear dynamics to model and control complex system" and from the European Union STREP project n. 012911 ``Human behavior through dynamics of complex social networks: an interdisciplinary approach". We would also like to thank Jean-Philippe Bouchaud, Szabolcs Mike, and Spyros Skouras for useful discussions, and Marcus Daniels and Claudia Coronello for technical support. }

\end{document}